\documentclass[onecolumn]{svjour3} 

\smartqed 
\PassOptionsToPackage{table}{xcolor}
\usepackage{tikz}
\usetikzlibrary{shapes}
\usepackage{graphicx}
\usepackage{booktabs}
\usepackage{algorithm,algpseudocode}
\usepackage{float}
\usepackage{amssymb}
\usepackage{amsmath}
\usepackage{bm}
\usepackage{hyperref}
\usepackage{natbib}
\usepackage{lineno}
\usepackage{soul}
\usepackage{placeins}
\usepackage{subcaption}

\usepackage{tabularx}
\usepackage[range-phrase = --,range-units = single, product-units = power, per-mode = symbol]{siunitx}
\DeclareSIUnit\pixel{px}
\DeclareSIUnit\voxel{vox}

\usepackage{multirow}
\usepackage{lineno}
\usepackage{comment}

\newcommand\Ha{\mbox{\text{Ha}}}
\DeclareMathOperator{\arccsc}{arccsc}

\begin{document}

\sloppy
% \title{A meshless approach to the super-resolution of turbulent statistics in 3D Ensemble PTV}
\title{Multi-objective optimization of  the magnetic wiping process in dip-coating}

\author{Fabio Pino$^{1,2}$ \and
        Benoit Scheid$^{2}$ \and
        Miguel A. Mendez$^{1}$
\thanks{Corresponding author: fabio.pino@vki.ac.be}}

\institute{   $^1$The von Karman Institute for Fluid Dynamics, EA Department, Waterloosesteenweg 72, Sint-Genesius-Rode 1640, Belgium\\
              $^2$Université libre de Bruxelles, Transfers, Interfaces and Processes (TIPs), Av. Franklin Roosevelt 50, Brussels 1050, Belgium}
\date{\today}
\maketitle
\abstract{
Electromagnetic wiping systems allow to pre-meter the coating thickness of the liquid metal on a moving substrate. These systems have the potential to provide more uniform coating and significantly higher production rates compared to pneumatic wiping, but they require substantially larger amounts of energy. This work presents a multi-objective optimization accounting for (1) maximal wiping efficiency (2) maximal smoothness of the wiping meniscus, and (3) minimal Joule heating. We present the Pareto front, identifying the best wiping conditions given a set of weights for the three competing objectives. The optimization was based on a 1D steady-state integral model, whose prediction scales according to the Hartmann number (Ha). The optimization uses a multi-gradient approach, with gradients computed with a combination of finite differences and variational methods.
The results show that the wiping efficiency depends solely on Ha and not the magnetic field distribution. Moreover, we show that the liquid thickness becomes insensitive to the intensity of the magnetic field above a certain threshold and that the current distribution (hence the Joule heating) is mildly affected by the magnetic field's intensity and shape. 
}

\section{Introduction}

Hot-dip galvanization consists in coating a steel substrate with a thin layer of zinc to prevent corrosion \citep{kistler1997coating}. The substrate usually takes the form of a large strip, dipped and then withdrawn from a bath of molten zinc. The liquid excess is removed via a wiping system. In galvannealing, the strip is heated for a few seconds after the wiping region in an induction furnace (annealing furnace) when the zinc is still liquid. Thereby, the iron in the steel and the zinc coating diffuse into each other, forming a zinc-iron inter-metallic compound with a composition of approximately $90\%$ iron and $10\%$ zinc in the bulk \citep{goodwin2013developments,galvingo_galvanneal}.

The wiping is usually performed through impinging gas jets in a process called jet wiping \citep{gosset2007jet,hocking2011deformations}. The process is simple and economically attractive, but it develops hydrodynamic instabilities when the substrate speed exceeds 2 or 3 m/s, depending on the wiping strength. These instabilities include splashing of the wiped liquid \citep{gosset2007jet} and oscillations of the impinging jet, which results in coating undulations \citep{Gosset2019,mendez2019experimental,Barreiro-Villaverde2021}.

An alternative to jet wiping is electromagnetic wiping, extensively investigated by \cite{european_union_report}. These systems are based on the ElectroMagnetic Brake (EMB) principle, that is the use of electromagnetic forces to control the thickness of the liquid coating \citep{ernst2009working,Lloyd-JonesCandBarkerHAandWorner1998d,Dumont2011d}. This is achieved using a strong magnetic field produced by a permanent magnet or high-frequency induction generators. The relative motion of the liquid film and the magnetic field create an induced current in the liquid film, which generates a downward Lorentz force that wipes the excess of liquid. These systems have been considered as ``pre-wiping" devices in lines that target very large substrate speeds \citep{Lloyd-JonesCandBarkerHAandWorner1998d}. 
Despite offering promising performances and coating uniformity at large production rates, the major challenges in their large-scale deployment are the energy requirements and the possible overheating of the liquid zinc, which increases the risks of promoting iron-zinc phase growth. The design of an electromagnetic system must thus strike a balance between the contrasting objectives of having sufficient ``wiping capabilities" while minimizing overheating and energy consumption.
 
This work investigates the trade-off between these competing goals using multi-objective optimization. Specifically, we present the Pareto front, defining the optimum according to three criteria: (1) wiping capabilities, (2) Joule heating and (3) smoothness of the wiping liquid meniscus. The optimization is based on a simple steady-state model, neglecting surface tension. The remaining of the article is structured as follows. Section \ref{integral_model} presents the simplified model used in the optimization; this is based on the steady-state streamwise momentum balance of the liquid film with the magnetic effect expressed by the Hartmann number. Section \ref{sens_derivatives} reports on the methods for computing the relevant first-order sensitivity of the problem, which combined variational methods and finite differences depending on the size of the parameter vector. Section \ref{Multiobjective_opt} defines the constrained multi-objective optimization problem and the solution method based on a multi-gradient descent. Finally, conclusions and perspectives are reported in Section \ref{conclusion}. 

\begin{figure}[h!]
    \centering
    \includegraphics[width=0.8\textwidth]{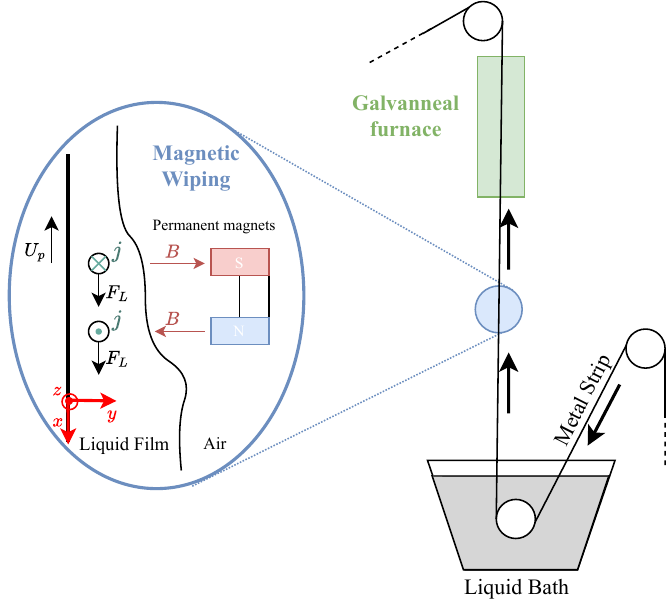}
    \caption{Schematic of the hot-dip galvannealed coating process with the metal strip dip into a bath of molten zinc, withdrawn vertically at a constant velocity $U_p$. Outside the bath, the liquid excess is wiped out by a magnetic actuator, and then the strip is heated in a galvannealing furnace.}
    \label{problem_scheme_1}
\end{figure}

\section{Problem description}
We consider liquid zinc with density $\rho=6570 $kg/m$^3$, dynamic viscosity $\mu=3.5\times 10^{-3} $ Pa$\cdot$s, kinematic viscosity $\nu=\mu/\rho$, surface tension $\sigma=0.7 $N/m and electrical conductivity $\sigma_M=0.028\times 10^{8}$ S/m. These properties are taken at around 500 Celsius degrees as in \cite{giordanengo2000new}. Figure \ref{problem_scheme_1} shows the schematic of the process: a metal strip is withdrawn from a bath of molten zinc at a constant speed $U_p$ under the effect of gravity $g$, wiped by permanent magnets (light blue circle) and then heated in a galvannealing furnace (green box). 

The reference system is centered on the substrate, with the positive streamwise axis $x$ pointing downwards, the wall-normal axis $y$ towards the free surface, and the cross-stream axis $z$ in the spanwise direction. We assume a two-dimensional velocity vector field within the liquid, given by $\mathbf{V}=[u(x,y),v(x,y),0]^T$. The flow is bounded by the substrate at $y = 0$, and by the free surface $h$, with the streamwise flow rate $q$ defined as:

\begin{equation}
\label{eq:flow_rate}
    q = \int_0^{h} u(y) \; dy.
\end{equation}
The magnetic field generated by the actuators is given by the vector:

\begin{equation}
    \mathbf{B}=[0,-B(x),0]^T,
\end{equation}
where $B(x)$ is always positive since, for simplicity, we are only considering the negative pole. The relative motion of the liquid zinc with respect to the external magnetic field generates an induced current in the bulk. Neglecting the electric potential difference, this is given by the vector:

\begin{equation}
    \mathbf{J} = \mathbf{V}\times\mathbf{B} = [0,0,j_z]^T = [0,0,-\sigma_M u B],
    \label{current_density}
\end{equation}
The integraction of the induced current with the external magnetic field results in a downward Lorentz force $\mathbf{F}_L$ defined as the cross product of $\mathbf{B}$ and $\mathbf{J}$:

\begin{equation}
    \mathbf{F}_L = \mathbf{J}\times\mathbf{B} = [F_L, 0, 0]^T,
\end{equation}
with $F_L = \sigma_M u B^2$. 
\FloatBarrier

%   --------------
%    Scaling quantities    
%   --------------
\section{Scaling quantities}
To recast the governing equations in nondimensional form, we scale the flow velocity by the substrate velocity $U_p$ and lengths by the liquid film thickness resulting from the steady state viscous-gravity balance \citep{mendez2021dynamics}:

\begin{equation}
    [u] = U_p \qquad\qquad [h] = \sqrt{\frac{\nu U_p}{g}}\,,
\end{equation}
where the $[\cdot]$ denotes reference quantities. Based on $[u]$ and $[h]$, we obtain the scale for the flow rate as:

\begin{equation}
    [q]=[u][h]=\sqrt{\frac{\nu U_p^3}{g}}
\end{equation}

We scale the streamwise coordinate $x$ assuming long-wave conditions:
\begin{equation}
    [x] = [h]/\epsilon,
\end{equation}
with $\epsilon\ll1.$

We define the magnetic field and the induced current reference quantities with their maximum values:
\begin{equation}
    [B]=sup(B(x)) \qquad\qquad [J]=\sigma_MU_p[B],
\end{equation}

We define the power density $P_D$ reference quantity as:
\begin{equation}
    [P_D] = U_p\rho g.
\end{equation}

The non-dimensional group governing the system is the Hartman number:

\begin{equation}
    \Ha = \sqrt{\frac{\sigma_M[B]^2U_p}{\rho g}},
\end{equation}

which weighs the importance of electromagnetic to gravitational forces. In the following, we use the hat notation $\hat{a}=a/[a]$ for the non-dimensional quantities and the shorthand notation $\partial_xh= \partial h/\partial_x$ for partial derivatives.

\section{Simplified model}
\label{integral_model}

In line with classical simplified engineering models for the jet wiping \citep{thornton1976analytical,tuck1983continuous}, we assume a gently sloped interface ($\partial_x h\sim 0$) \citep{mendez2021dynamics}, a unidimensional flow ($v(x,y)=0$), without surface tension \citep{Dumont2011d} and without pressure gradient along the wall-normal and the streamwise direction. The governing equation is given by the steady-state streamwise momentum balance, which in non-dimensional form reads:

\begin{equation}
    \label{gover_equation}
   \partial_{\hat{y}\hat{y}}\hat{u} + 1 - \Ha^{2}\hat{B}^2\hat{u} = 0,
\end{equation} where the first term is the viscous stress, the second the gravitational forces, and the last the Lorentz force. The non-dimensional magnetic field $\hat{B}$ is modelled as a Gaussian along the streamwise direction, centred in $\hat{x}=0$ with standard deviation $\gamma$:

\begin{equation}
    \hat{B} = \exp\Big(-\frac{\hat{x}^2}{2\gamma^2} \Big).
\end{equation}

We impose a no-slip boundary condition at the wall, and a shear-free boundary condition at the free surface \cite[Chapter~2]{kalliadasis2011falling}:

\begin{equation}
    \label{adim_bcs}
    \hat{u}(\hat{y}=0) = -1, \qquad\qquad  \partial_{\hat{y}}\hat{u}(\hat{y}=\hat{h})=0.
\end{equation}

The solutions of \eqref{gover_equation} with boundary conditions \eqref{adim_bcs} is given by:

\begin{equation}
\begin{split}
\label{final_eq_velocity}
    \hat{u} =  \left(1 + \frac{1}{\Ha^2\hat{B}^2}\right)&  \Big(\tanh{(\Ha\hat{h}\hat{B})}\sinh{(\Ha\hat{y}\hat{B})} - \cosh{(\Ha\hat{y}\hat{B})}\Big) + \frac{1}{\Ha^2\hat{B}^2},
\end{split}
\end{equation} as shown by \cite{Dumont2011d}. Integrating the streamwise velocity over the liquid film thickness, we obtain a relation between $\hat{q}$ and $\hat{h}$:

\begin{equation}
\begin{split}
    \label{final_q_func_2}
    \hat{q} = g(\hat{h};\Ha\hat{B}) = &\frac{1}{\Ha^3\hat{B}^{3}} \Big\{ \Ha\,,\hat{h}\hat{B} - \tanh{(\Ha\,\hat{h}\hat{B})}\left(\Ha^2\hat{B}^2+1\right)\Big\}.
\end{split}
\end{equation}
\begin{figure}[h!]
\centering
% For eps or pdf graphics
\includegraphics[width=0.8\textwidth]{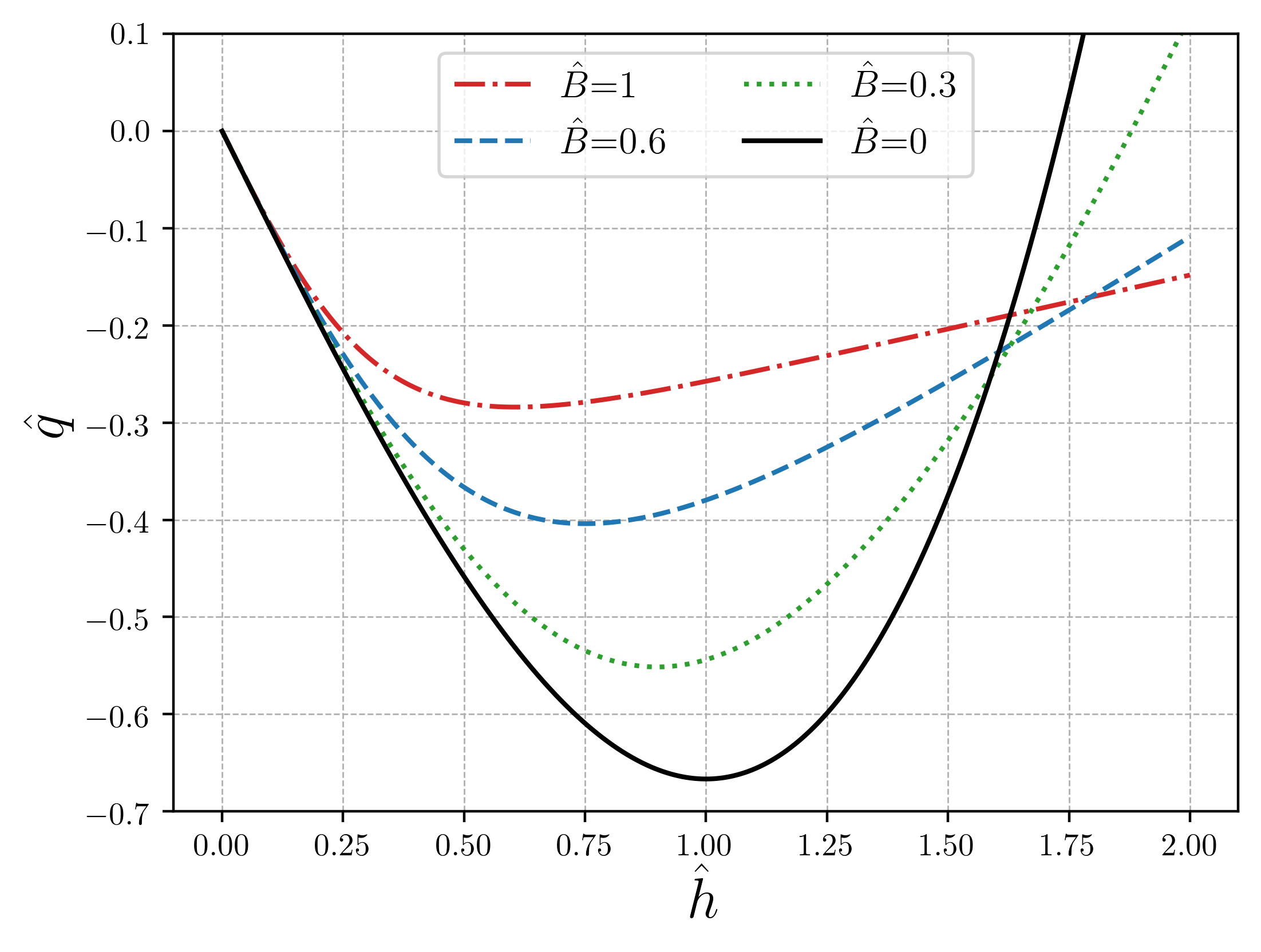}
%  \raisebox{1.3in}{a)}\input{figa}\quad
%  \raisebox{1.3in}{b)}\input{figb}
\caption{Relation between the nondimensional flow rate $\hat{q}$ and liquid film thickness $\hat{h}$ for different value of the nondimensional magnetic field $\hat{B}$.}
\label{relation_h_q_diff_B}
\end{figure}

For $\Ha\rightarrow0$, \eqref{final_eq_velocity} and \eqref{final_q_func_2} retrieve the relations without wiping \citep{mendez2021dynamics}. Figure \ref{relation_h_q_diff_B} plots \eqref{final_q_func_2} for a fixed $\Ha$ at a fixed $\hat{x}$ location for different values of the magnetic field $\hat{B}$. In wiping conditions, $\hat{B}$ is varying along $\hat{x}$ with a minimum $\hat{B}=0$ far from the actuator and a maximum $\hat{B}=1$ right at the pole location ($\hat{x}=0$). The thickness $\hat{h}$ reaches a minimum at $\hat{x}=0$. Its value is found by solving $\partial_{\hat{h}}g(\hat{h};\Ha,\hat{B}=1)=0$, which yields:

\begin{equation}
\label{eq:h_min}
    \hat{h}_{min}=\frac{1}{\Ha}\arccsc\Big(\sqrt{\frac{1}{\Ha^2 + 1}}\Big).
\end{equation}

The flow rate $\hat{q}_{opt}$ is obtained injecting \eqref{eq:h_min} in \eqref{final_q_func_2}. The liquid film profile $\hat{h}(\hat{x})$ is given by solving the equation:

\begin{equation}
    \hat{q}_{opt} - g(\hat{h}(\hat{x});\Ha,\hat{B}(\hat{x})) = 0,
\end{equation}
at different locations along the stream-wise coordinate $\hat{x}$. In this work, we consider $N=100$ equispaced points in a domain $\hat{x}\in[-10,10]$.

%   --------------
%    Sensitivity Analysis    
%   --------------
\section{Sensitivity derivatives}
\label{sens_derivatives}

The gradient-based multiobjective optimization carried out in this work requires the computation of the jacobian (variation) of the state variables $\hat{h}$, $\hat{q}$, $\hat{u}(\hat{y})$ and $\hat{j}$ with respect to the magnetic wiping parameters $\mathbf{p}=[\Ha,\hat{B}]^T$:

\begin{equation}
    \partial_{\mathbf{p}}\hat{h}(\hat{x}), \qquad\qquad
    \partial_{\mathbf{p}}\hat{u}(\hat{y}), \qquad\qquad
    \partial_{\mathbf{p}}\hat{q}, \qquad\qquad
    \partial_{\mathbf{p}}\hat{j}(\hat{x},\hat{y})\,.
\end{equation}

For a given $\Ha$ and $\hat{B}(x)$, the gradient $\partial_{\mathbf{p}}\hat{h}$ was computed with finite differences using the python library \textit{scipy.optimize.approx\_fprime} on a set of $N$ equispaced points:

\begin{equation}
    \hat{h}(\hat{x}_i) = \frac{\hat{h}(\hat{x}_i; \mathbf{p}+\delta) - \hat{h}(\hat{x}_i; \mathbf{p})}{\delta},
\end{equation} 
with $\delta=10^{-8}$.

The total variation of $\hat{u}$ ($d\hat{u}/d\mathbf{p}$) is given by:

\begin{equation}
    \frac{d\hat{u}}{d\mathbf{p}} = \partial_{\mathbf{p}}\hat{u} + \partial_{\hat{h}}\hat{u}\partial_{\mathbf{p}}\hat{h} \,,
\end{equation} where $\partial_{\hat{h}}\hat{u}$ was obtained differentiating \eqref{final_eq_velocity}:

\begin{equation}
    \partial_{\hat{h}}\hat{u} = \left(1 + \frac{1}{\Ha^2\hat{B}^2}\right)\frac{\Ha\hat{B}}{\cosh{(\Ha\hat{h}\hat{B})}^2}\sinh{(\Ha\hat{y}\hat{B})} \,.
\end{equation}

The sensitivity gradient $\partial_{\mathbf{p}}\hat{u}(\hat{y})$ was calculated using a variational method \citep{calver2017numerical}. This consists in differentiating \eqref{gover_equation} with respect to $\mathbf{p}$, which leads to a system for the sensitive gradients given by:

\begin{align}
    \label{variationa_approach}
    \partial_{\mathbf{p}}\frac{d\mathbf{u}(\hat{y})}{d\hat{y}}\, =  \, \mathbf{F}_{\mathbf{u}}\mathbf{S}_{\mathbf{p}} + \mathbf{F}_{\mathbf{p}}\,,
\end{align} where $\mathbf{u} = [u_1,u_2]^T = [\hat{u}(\hat{y}), \partial_{\hat{y}}\hat{u}(\hat{y})]^T$, and  $\mathbf{F}_{\mathbf{u}}$ and $\mathbf{F}_{\mathbf{p}}$ are the Jacobian matrices with respect to $\mathbf{u}$ and $\mathbf{p}$ respectively:
\begin{equation}
    \mathbf{F_u} = \begin{pmatrix}
0 & 1\\
\Ha^2\hat{B}^2 & 0
\end{pmatrix}\, \qquad
\mathbf{F}_{\mathbf{p}} = \begin{pmatrix}
0 & 0\\
2\Ha\hat{B}^2u_1  & 2\Ha^2\hat{B}u_1
\end{pmatrix},
\end{equation} and $\mathbf{S_p}$ is the matrix of sensitivity derivatives of $u_1$ and $u_2$:

\begin{equation}
    \mathbf{S_p} = \begin{pmatrix}
\partial_{\Ha}u_{1} & \partial_{\hat{B}}u_{1}\\
\partial_{\Ha}u_{2}  & \partial_{\hat{B}}u_{2}.
\end{pmatrix}.
\end{equation}

Solving \eqref{gover_equation} and \eqref{variationa_approach} with boundary conditions: 

\begin{equation}
    \begin{split}
         u_1(\hat{y})|_{\hat{y}=0} = -1, \quad u_2(\hat{y})|_{\hat{y}=\hat{h}}, = 0,\quad
        \partial_{\Ha}u_{1}|_{\hat{y}=0} = 0,\\ \partial_{\hat{B}}u_{1}|_{\hat{y}=0} = 0,\quad
        \partial_{\Ha}u_{2}|_{\hat{y}=\hat{h}} = 0, \quad \partial_{\hat{B}}u_{2}|_{\hat{y}=\hat{h}} = 0.     
    \end{split}
\end{equation} gives the vector of sensitivity derivatives $\partial_{\mathbf{p}}\hat{u} = [\partial_{\Ha}\hat{u}, \partial_{\hat{B}}\hat{u}]^T$.

The variation of the flow rate $\partial_{\mathbf{p}}\hat{q}$ is obtained as: 

\begin{equation}
\label{eq:gradient_flow_rate}
    \frac{d\mathbf{q}}{d\mathbf{p}} = \frac{d}{d\mathbf{p}}\int_{0}^{\mathbf{h}}\,\hat{u}(\hat{y}')\,d\hat{y}' = \hat{u}(\hat{h})\partial_{\mathbf{p}}\hat{h} + \int_{0}^{\hat{h}}\,\partial_{\mathbf{p}}\hat{u}(\hat{y}')\,dy'.
\end{equation}

Based on its definition in \eqref{current_density}, the current density sensitivities are given by: 

\begin{equation}
    \partial_{\Ha}\hat{j} = -\partial_{\Ha}\hat{u}(\hat{y})\hat{B}(\hat{x}),\qquad\qquad \partial_{\hat{B}}\hat{j} = -\hat{u}(\hat{y}) -  \partial_{\hat{B}}\hat{u}(\hat{y})\hat{B}(\hat{x}).
\end{equation}

% Multiobjective optimization
\section{Multiobjectives optimization}
\label{Multiobjective_opt}

We set up a constrained multiobjectives minimization problem problem for the parameters $\Ha$ and $\gamma$, controlling the intensity and the shape of the magnetic field distribution. The constraints impose that $\Ha <10$, above which the magnetic field becomes generally overly large, a positive standard deviation $\gamma>0$, and a negative flow rate $\hat{q}<0$ (that is some liquid is withdrawn from the plate):

\begin{equation}
\label{eq:opt_problem}
\boxed{
\begin{aligned}
\min_{\Ha,\gamma} \quad & J(\Ha,\gamma)\\
\textrm{s.t.} & \gamma>0, \,\,\Ha<10 \,\,\hat{q}<0    \\
\end{aligned}}
\end{equation}

The cost function $J$ adds up the contribution of three terms, weighting the wiping strength ($J_1$), the smoothness of the free surface ($J_2$) and the Joule heating due to the induced currents ($J_3$):

\begin{equation}
\label{functional_opt}
    J(Ha,\gamma) = J_1 + J_2 + J_3\,,
\end{equation} where:

\begin{equation}
    \begin{split}
    J_1 = -\hat{q},\quad
    J_2 = \frac{1}{2}\int\limits_{x_{1}}\limits^{x_{2}}\Big(\partial_{\hat{x}\hat{x}}\hat{h}(\hat{x})\Big)^2\;d\hat{x},\quad
    J_3 = \frac{1}{2}\int\limits_{x_{1}}\limits^{x_{2}}\int\limits_{0}\limits^{\hat{h}}\Ha^2\hat{j}(\hat{x},\hat{y})^2\;d\hat{y}d\hat{x}\,.  
    \end{split}
\end{equation}

Reducing the term $J_1$ requires reducing the flow rate and hence the liquid film thickness. Reducing the term $J_2$ requires reducing the curvature of the wiping meniscus, hence producing smoother transition from the thin film solution at $x\rightarrow -\infty$ and the thick solution at $x\rightarrow \infty$. A smoother transition implies smoother curvature of the flow streamlines within the film, smaller velocity gradients and thus a more stable film. The second derivative was computed by differentiating a fourth-order spline, which interpolated the solution in all the grid points.  

The term $J_3$ is related to the Joule heating due to the induced current $\mathbf{J}$. In its differential form, the heating power $P_D$ is equal to the dot product of the current density vector $\mathbf{{J}}$ times the electric field $\mathbf{E}$. Since the current density and the electric field are linked by the relation $\mathbf{E}=\mathbf{J}\sigma_M$, this yields the following expression for $P_D$:

\begin{equation}
    P_D = \mathbf{J}\cdot\mathbf{E} = \mathbf{J}\cdot\mathbf{J}/\sigma_M = ||\mathbf{J}||^2/\sigma_M\,,
    \label{eq_power_density_1}
\end{equation} which in non-dimensional form reads (see also \cite{smolentsev1997laminar} and \cite{mao2008joule}):

\begin{equation}
    \hat{P}_D = \frac{\sigma_MU_p^2[B]^2}{[P_D]}\hat{j}^2 =  \Ha^2\hat{j}^2.
\end{equation}

\subsection{Multi-Gradient Descent method}
The constrained optimization problem \eqref{eq:opt_problem} is solved using the Stochastic Multi-Subgradient Descent Algorithm (SMSGDA) method \citep{milojkovic2019multi}. This consists in updating the value of the parameters, starting from an initial value $\mathbf{p}_{init}$, in a gradient descent fashion:

\begin{equation}
    \mathbf{p}^{\{i+1\}} = \mathbf{p}^{\{i\}} - \eta\nabla_{\mathbf{p}}J(\mathbf{p}) 
\end{equation} where the superscript$^{\{i\}}$ indicates the number of optimization iterations, $\eta$ is a small number controlling the parameter update and $\nabla_{\mathbf{p}}J(\mathbf{p})$ is the common descent vector computed as described in the following. This vector is given by the linear combination of the gradient of the ith objective function  $\nabla_{\mathbf{p}} J_i(\mathbf{p})=[\partial_{\Ha}J_i,\partial_{\gamma}J_i]^T$ (see \ref{sec:gradients_computation} for more details):

\begin{equation}
    \nabla_{\mathbf{p}}J(\mathbf{p}) = \sum_{i=1}^3\,\alpha_i\nabla_{\mathbf{p}} J_i(\mathbf{p}),
\end{equation}
where $\alpha_i$ are the weighting coefficients. The values of these weights are obtained by solving a quadratic optimization problem at each optimization iteration: 

\begin{equation}
    \min_{\alpha, \cdots,\alpha_n} \quad \Big[||\sum_{i=1}^n\,\alpha_i\nabla_{\mathbf{p}}J_{i}(\mathbf{p})||^2\;|\sum_{i=1}^n\,\alpha_i = 1,\alpha_i \geq 0\Big] \\
\end{equation} where $||\cdot||$ is the Euclidean norm.
Due to the different scales of the objective functions, the gradients can have significantly different norms along the two objectives, hence causing these two to dominate on the other one. To mitigate this risk, we normalize the gradients as follows:

\begin{equation}
    \nabla_{\mathbf{p}}\hat{J}_i(\mathbf{p}) = \frac{\nabla_{\mathbf{p}}J_i(\mathbf{p})}{J_i(\mathbf{p}_{init})}\,,
\end{equation} where $\nabla_{\mathbf{p}}\hat{J}_i(\mathbf{p})$ is the normalized gradient vector, $\nabla_{\mathbf{p}}J_i(\mathbf{p})$ is the non-normalized gradient and $J_i(\mathbf{p}_{init})$ is the initial loss related to the initial parameter guess $\mathbf{p}_{init}$. 

The optimization is concluded when the necessary optimality Karush-Kuhn-Tucker (KKT) condition is met. This reads:

\begin{equation*}
    \alpha,\cdots,\alpha_n \geq 0\qquad, \sum_{i=1}^n\,\alpha_i = 1\qquad,
      \sum_{i=1}^n\,\alpha_i\nabla_{\mathbf{p}}J_i(\mathbf{p})=0\,.
\end{equation*}
Every solution satisfying these conditions is called Pareto Stationary \citep{desideri2012multiple}. To find the Pareto optimal solutions, we run several simulations starting from different $\mathbf{p}_{init}$ until they converged on the point of the Pareto front.

\section{Results}
This section reports the sensitivity derivatives (Subsection~\ref{subsec:sens_der}) and the Pareto front resulting from the multi-objective optimization problem in the parameter $\Ha-\gamma$ and the objective functions $J_1-J_2-J_3$ spaces (Subsection~\ref{subsec:multiobgopt}).

\subsection{Sensitivity derivatives}
\label{subsec:sens_der}
\begin{figure}
\centering
% For eps or pdf graphics
\includegraphics[width=0.6\linewidth]{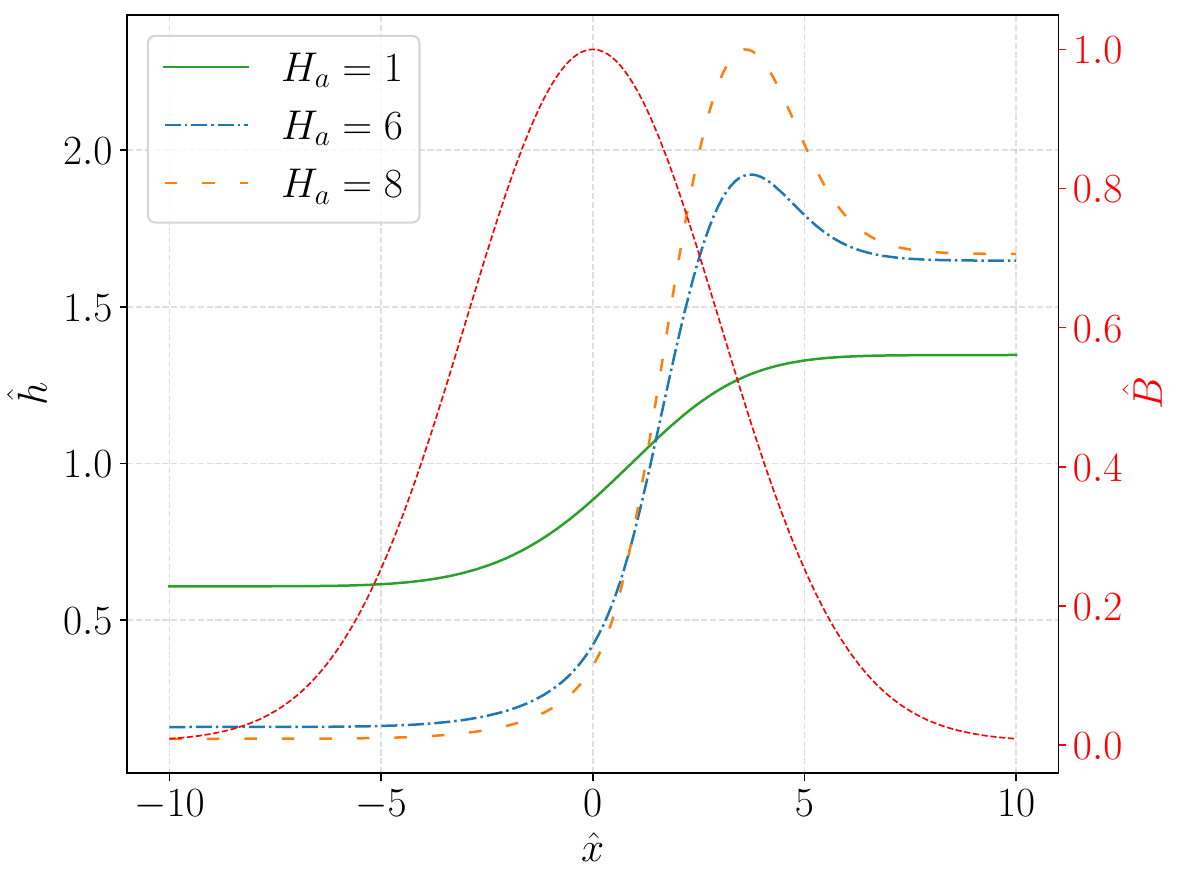}
%  \raisebox{1.3in}{a)}\input{figa}\quad
%  \raisebox{1.3in}{b)}\input{figb}
\caption{\small Examples of steady-state solution for different intensities of the magnetic fields with the appearance of a hump, for different values of the Hartman number $\Ha$.}
\label{hump_formation}
\end{figure}

Figure~\ref{hump_formation} shows the liquid film for a magnetic Gaussian with $\gamma=3$ for different $\Ha$. As $\Ha$ increases, a portion of the film upstream of the actuator curves, creating a local peak that becomes higher and steeper as $\Ha$ increases. A hump starts to appear for $\Ha$ larger than two, approximately. The velocity field beneath this hump is further analyzed in Subsection \ref{subsec:multiobgopt}, where it is illustrated for some of the optimal solutions belonging to the Pareto front.

% Liquid film thickness
\subsubsection{Liquid film thickness}
\begin{figure}[!h]
\centering
  \begin{subfigure}[b]{0.49\textwidth}
    \includegraphics[width=\textwidth]{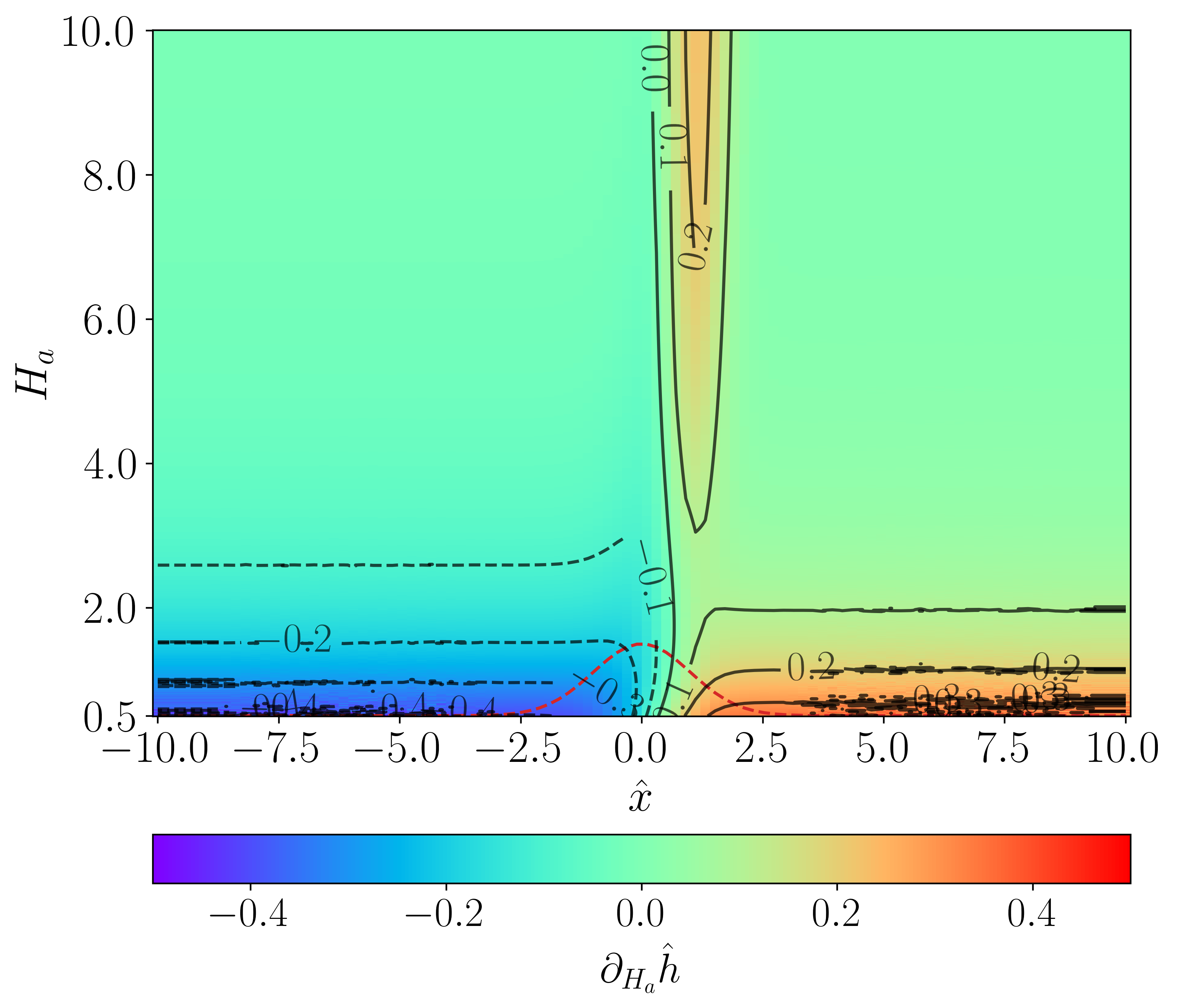}
    \caption{}
  \end{subfigure}
  \hfill
  \begin{subfigure}[b]{0.49\textwidth}
    \includegraphics[width=\textwidth]{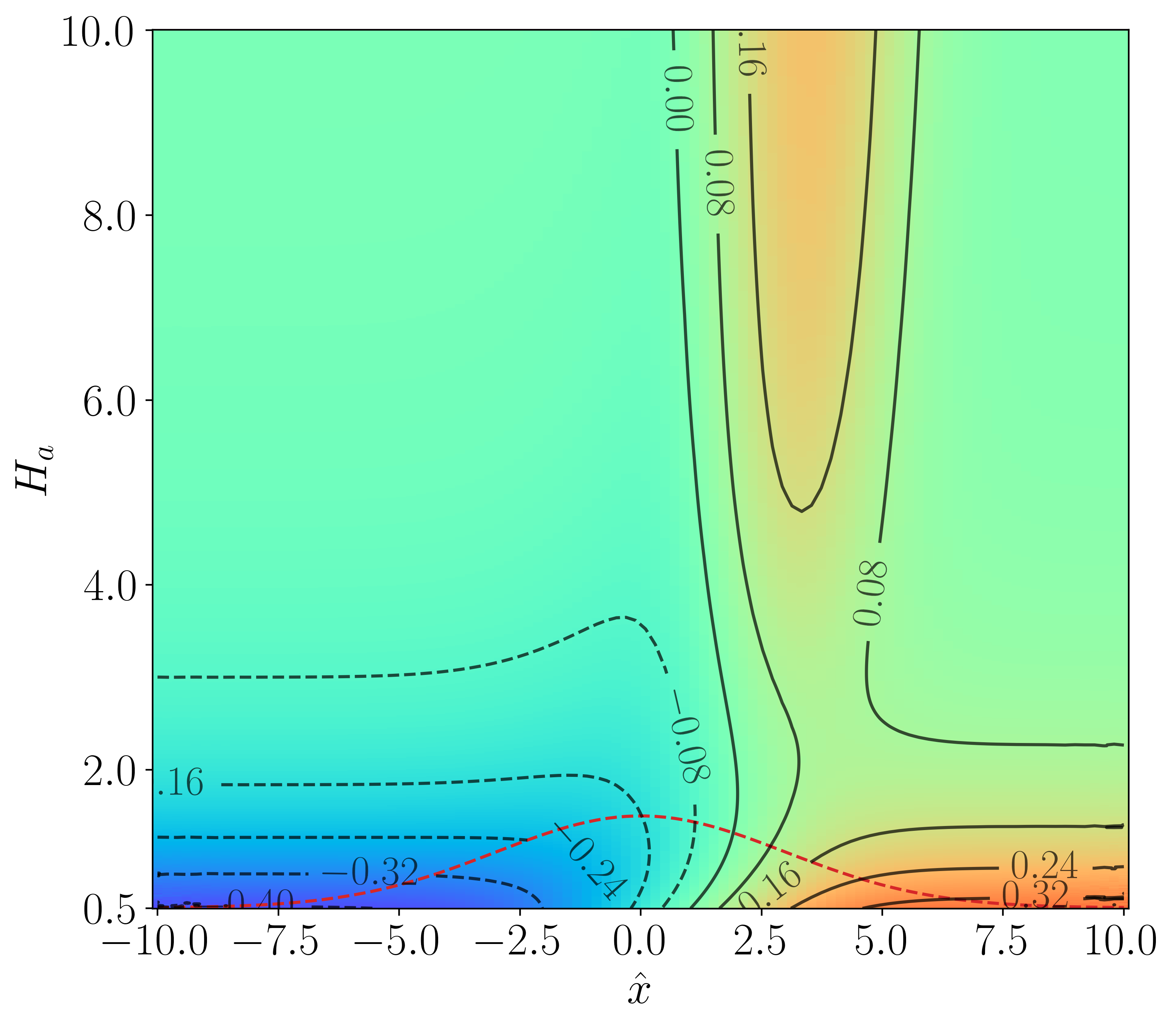}
    \caption{}
  \end{subfigure}
  \caption{Colour map of the liquid film height sensitivity with respect to the Hartmann number $Ha$ for a Gaussian magnetic field centred in zero (red dashed line) with a standard deviation of (a) 1 and (b) 3. The horizontal axis shows the one-dimensional streamwise coordinate $\hat{x}$, and the vertical axis shows the different values of $H_a$ and the colours show the gradient.}
  \label{sensitivity_h_Ha}
\end{figure}

Figure~\ref{sensitivity_h_Ha} shows the sensitivity of the film thickness $\hat{h}$ with respect to the Hartmann number $\Ha$ for two different magnetic field standard deviations: $\gamma=1$ and $\gamma=3$. In both cases, we remark that $\hat{h}$ becomes insensitive to the magnetic field for $\Ha>4$, apart from a small region of high sensitivity where a stationary hump starts to form. In addition, we notice that as $\gamma$ grows, the hump start to form at a higher value of $\Ha$. The sensitivity presents a non-symmetric behaviour around the symmetry axis of the magnetic field ($\hat{x}=0$). It has a larger negative region that passes over the centreline of the magnetic Gaussian, affecting points of the wiping region closer to the ran-back. A smaller $\gamma$ leads to the same qualitative behaviour but with steeper gradients. 
\FloatBarrier

% Induced current
\subsubsection{Induced current density}
\begin{figure*}[h!]
\centering
  \begin{subfigure}[b]{0.49\textwidth}
    \includegraphics[width=\textwidth]{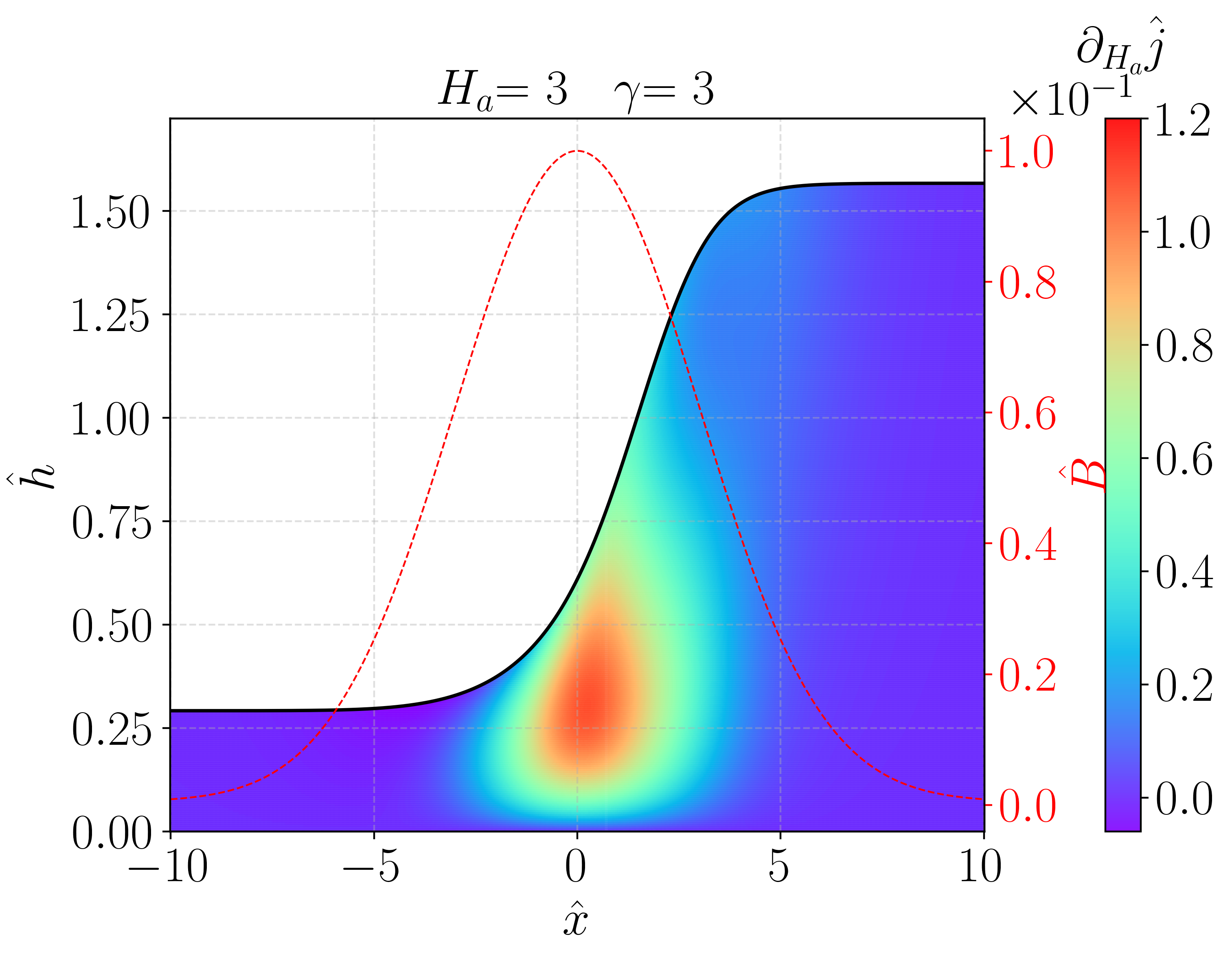}
    \caption{}
  \end{subfigure}
  \hfill
  \begin{subfigure}[b]{0.49\textwidth}
    \includegraphics[width=\textwidth]{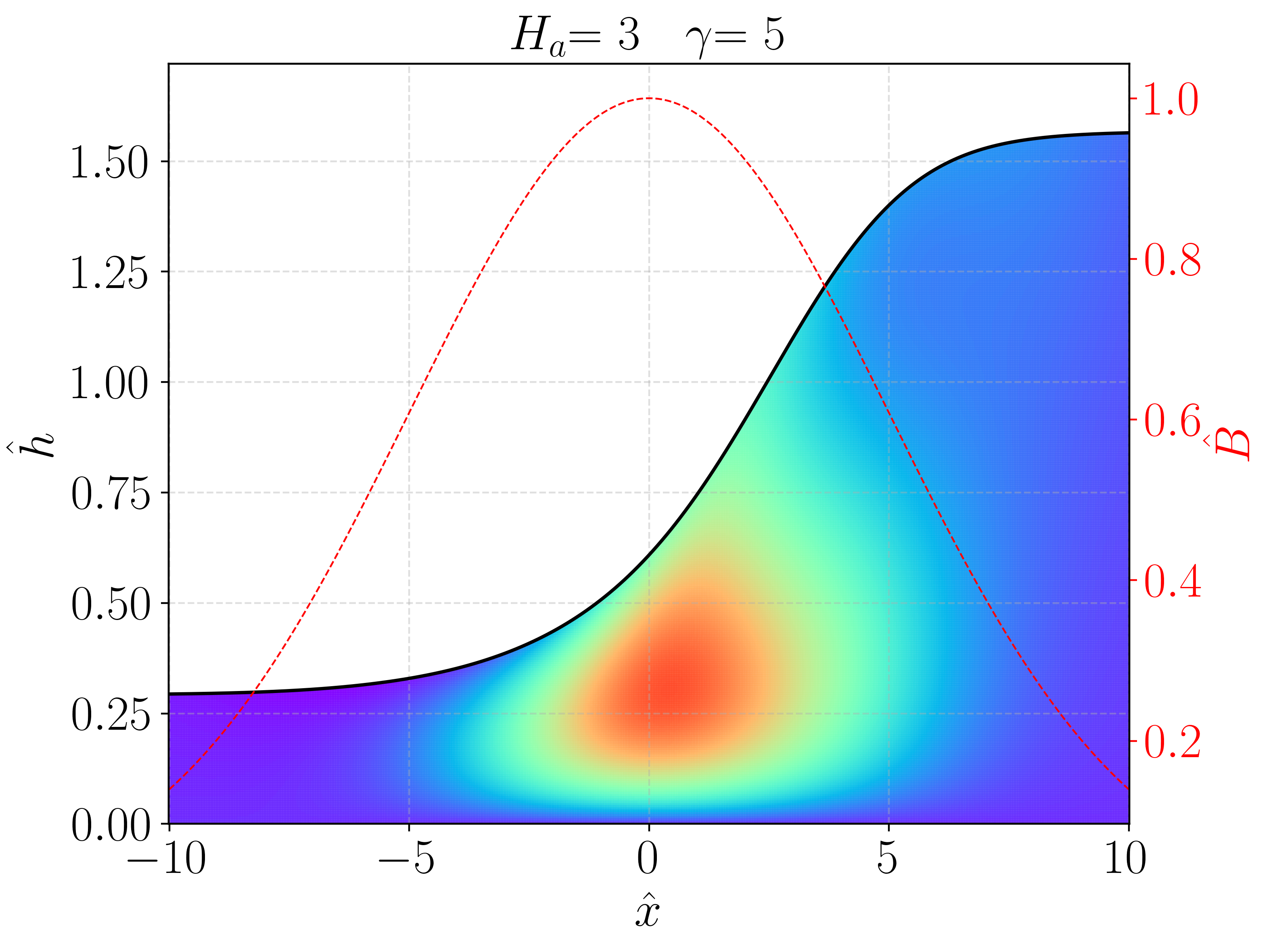}
    \caption{}
  \end{subfigure}
  \hfill
  \begin{subfigure}[b]{0.49\textwidth}
    \includegraphics[width=\textwidth]{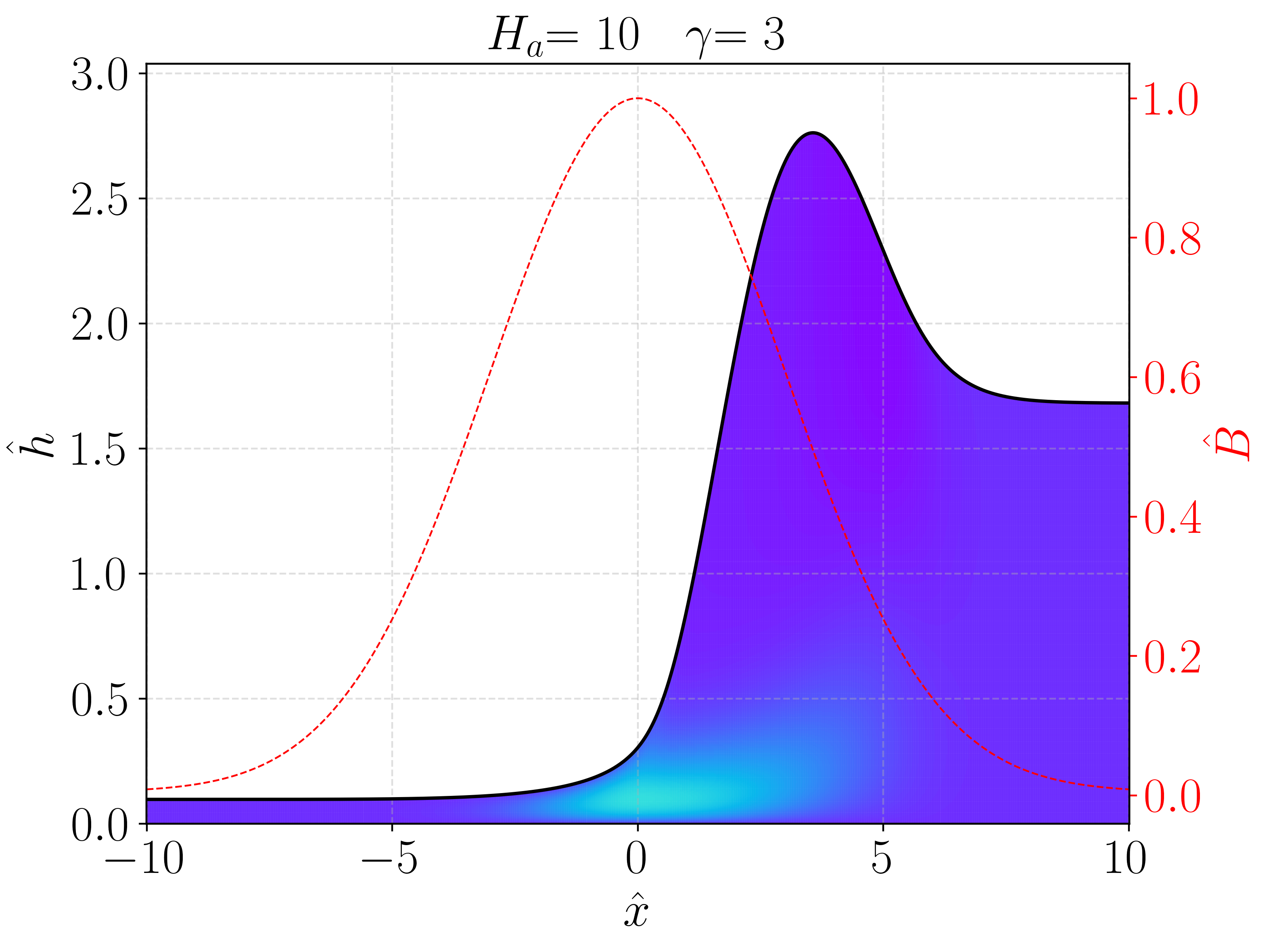}
    \caption{}
  \end{subfigure}
  \hfill
  \begin{subfigure}[b]{0.49\textwidth}
    \includegraphics[width=\textwidth]{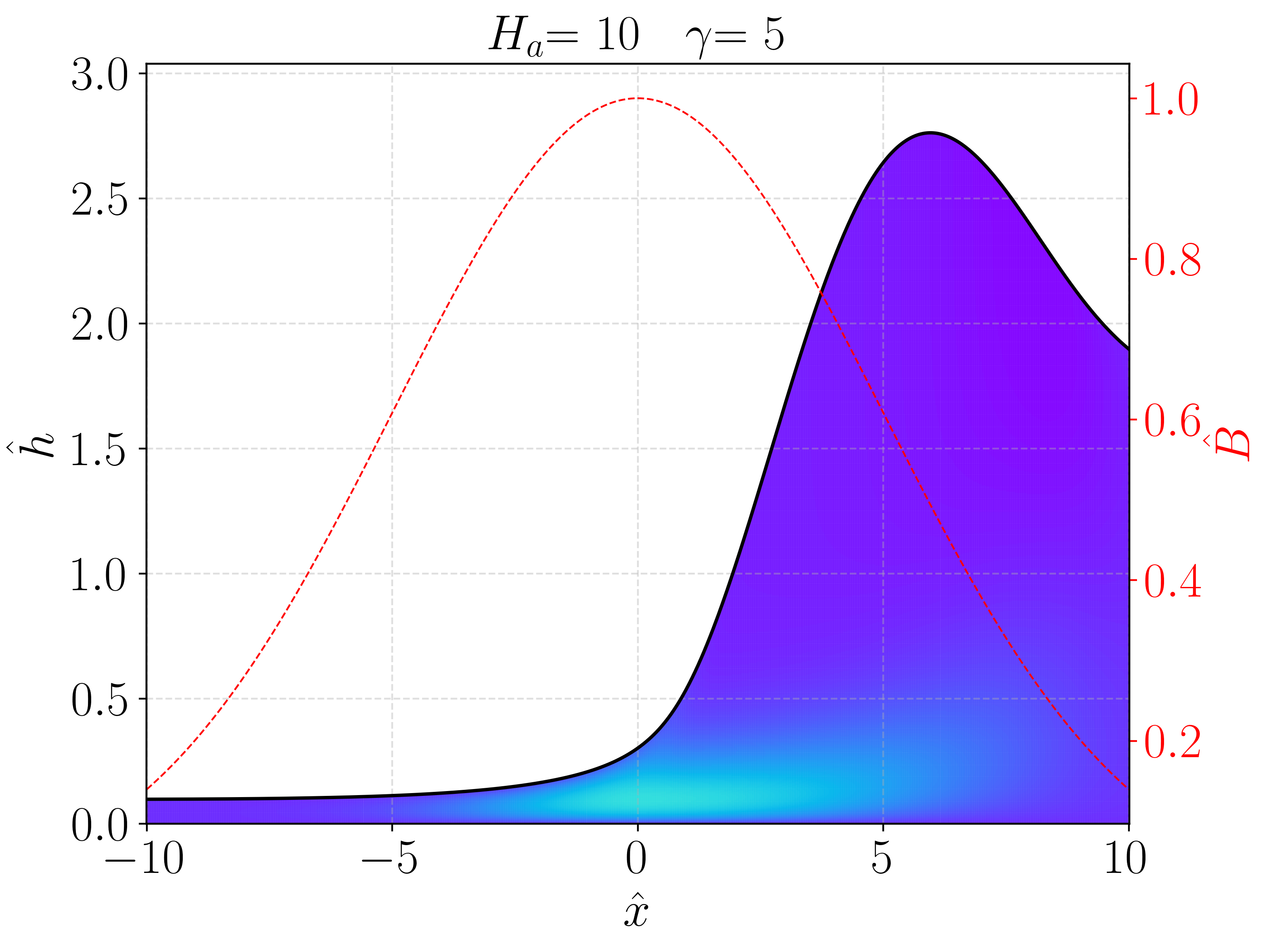}
    \caption{}
  \end{subfigure}
  \caption{Colour map of the induced current sensitivity sensitivities for Hartmann number (a and b) $Ha=3$ and (c and d) $\Ha=10$ and for magnetic field standard deviations (a and c) $\gamma=3$ and (b and d) $\gamma=5$.}
  \label{current_senstitivity_Ha}
\end{figure*}

Figure~\ref{current_senstitivity_Ha} shows the sensitivity of the electric current density function $j(\hat{x},\hat{y})$ with respect to the Hartmann number at (a and b) $Ha=3$ and (c and d) $Ha=10$ and with respect to the magnetic Gaussian standard deviations (a and c) $\gamma=3$ and (b and d) $\gamma=5$. For small $\Ha$, the most sensitive region is close to the peak of the magnetic Gaussian at the middle of the liquid film. The region of high sensitivity extends around this maximum, mainly in the run-back region of the film. It is interesting to note that the most sensitive area in this region is close to the free surface. In the far-field part, the sensitivity has a negative sign and is very small in magnitude. As for the thickness sensitivity, an increase of $\Ha$ makes the current density's sensitivity very small in the whole domain apart from a small region close to the magnetic peak and in the run-back.  

\FloatBarrier
% Optimization
\subsection{Multi-objectives optimization}
\label{subsec:multiobgopt}
\begin{figure}
     \centering
     \begin{subfigure}[b]{0.48\textwidth}
         \centering
         \includegraphics[width=\textwidth]{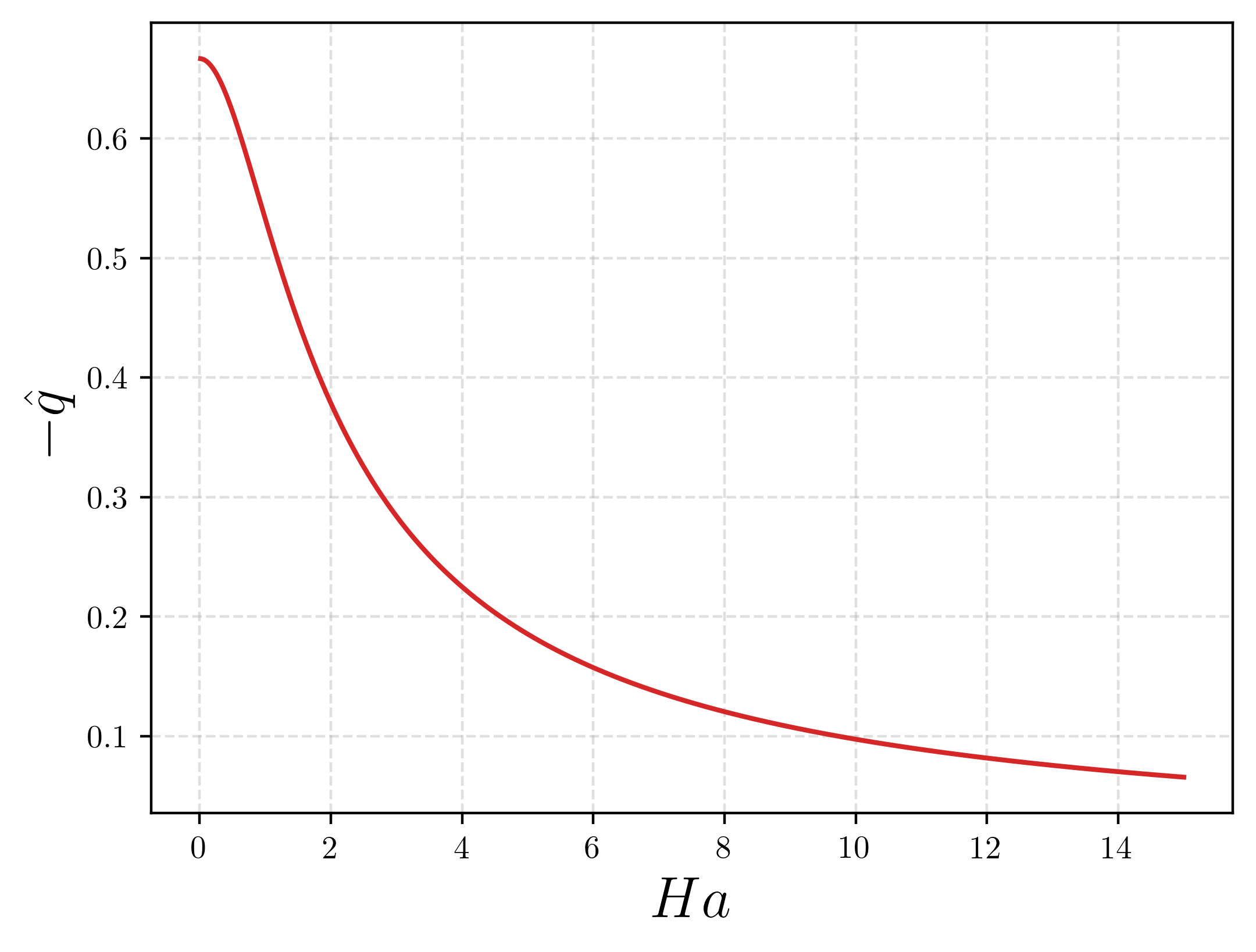}
         \caption{}
         \label{}
     \end{subfigure}
     \hfill
     \begin{subfigure}[b]{0.48\textwidth}
         \centering
         \includegraphics[width=\textwidth]{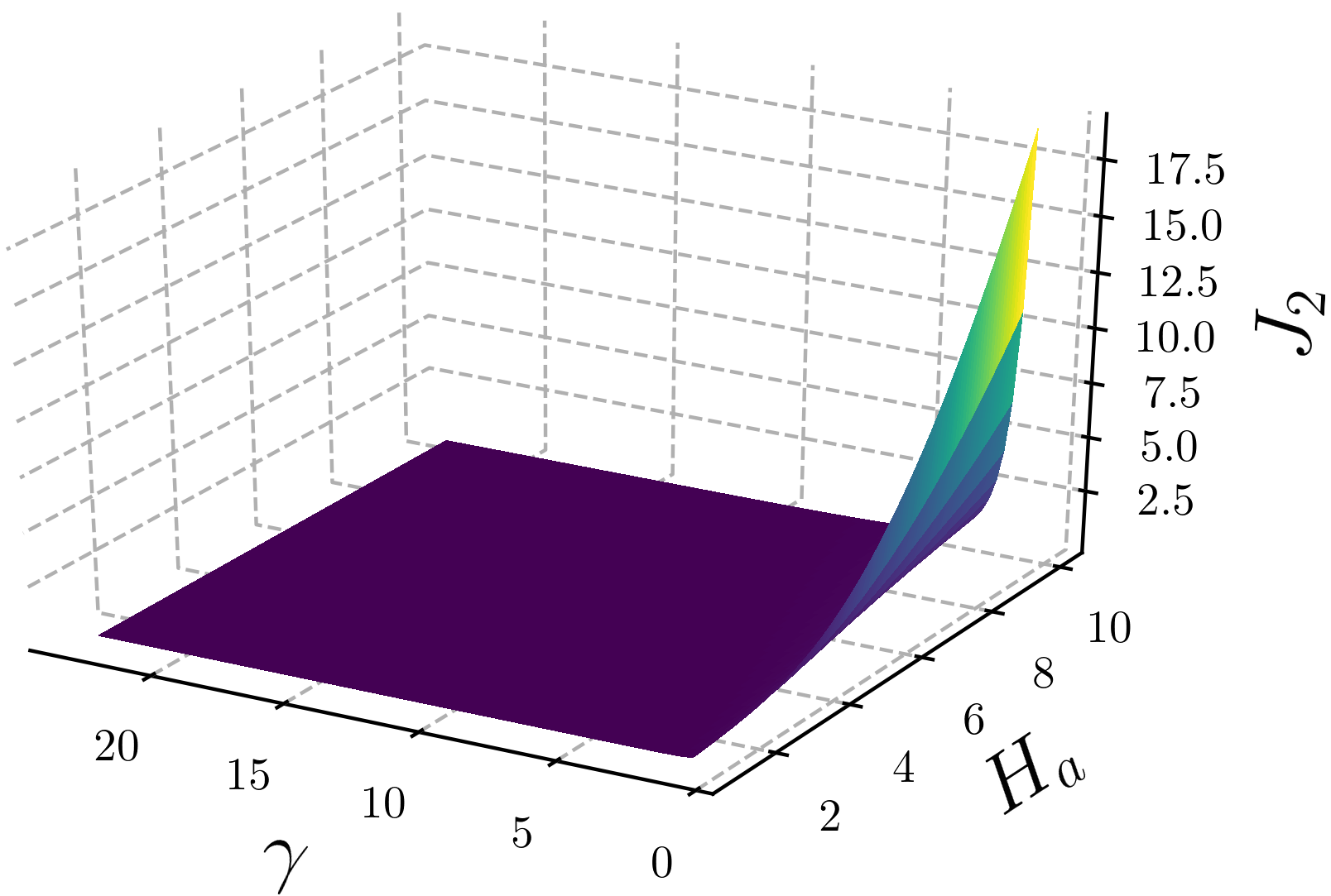}
         \caption{}
         \label{}
     \end{subfigure}
     \hfill
     \begin{subfigure}[b]{0.48\textwidth}
         \centering
         \includegraphics[width=\textwidth]{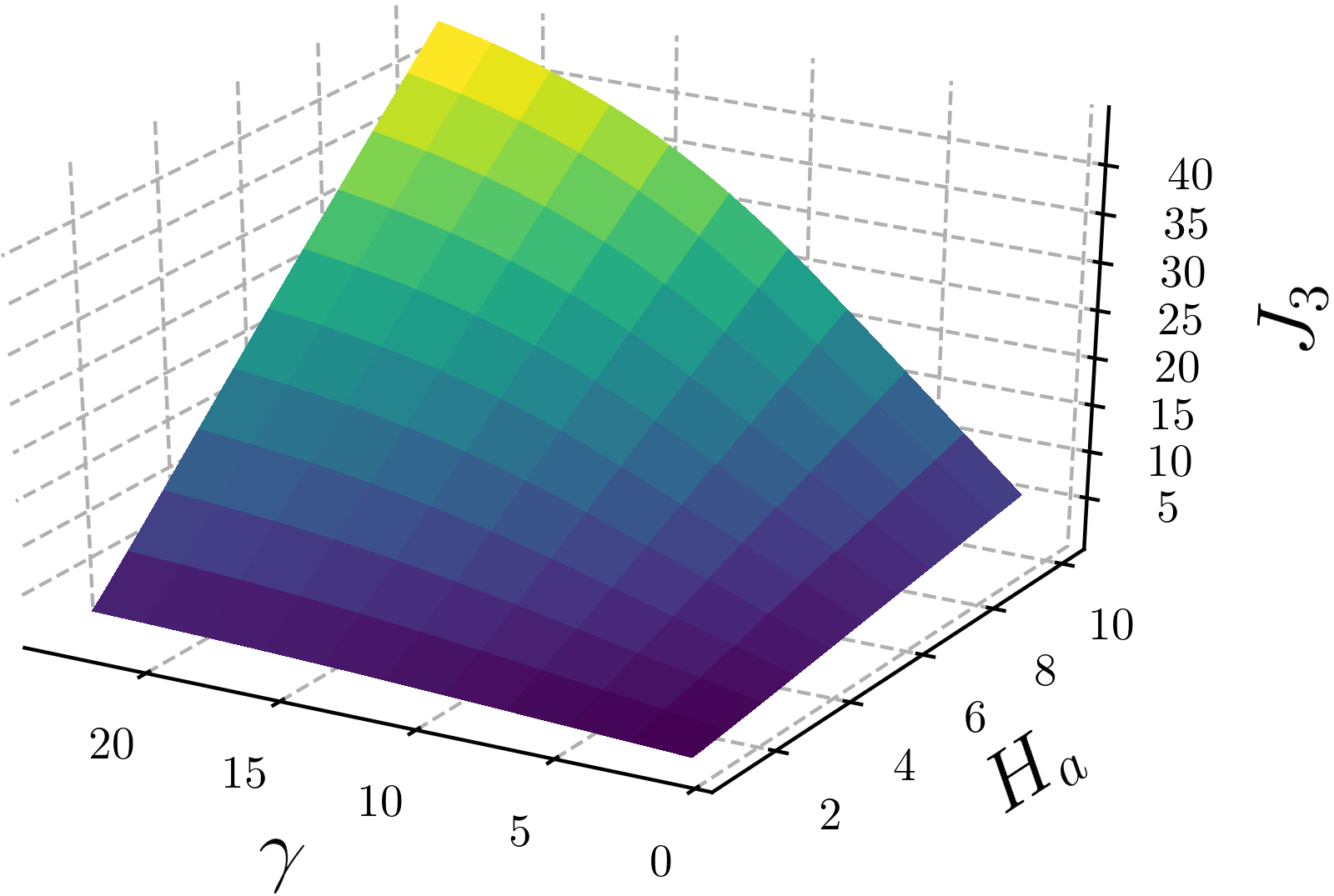}
         \caption{}
         \label{}
     \end{subfigure}
        \caption{Plot of the non-dimensional flow rate $\hat{q}$ the Joule heating (b) and of the liquid film curvature (c) as a function of the Hartmann number $H_a$ and the magnetic field standard deviation $\gamma$}
        \label{joule_heating_vs_Ha_epsilon}
\end{figure}

Figure~\ref{joule_heating_vs_Ha_epsilon} show $J_1$, $J_2$ and $J_3$ as a function of $Ha$ and $\gamma$. We notice their evident conflicting behavior from their different shapes. The curvature term $J_2$ tends to increase with $\Ha$ and decrease with $\gamma$, peaking at $\Ha=10$. The joule heating term $J_3$ grows with $\Ha$ and $\gamma$. As seen in Subsec.\ref{integral_model}, the flow rate depends solely on $\Ha$ with a monotone behavior that tends to become less steep for $\Ha>4$.

\begin{figure}[h]
\centering
  \begin{subfigure}[b]{0.49\textwidth}
    \includegraphics[width=\textwidth]{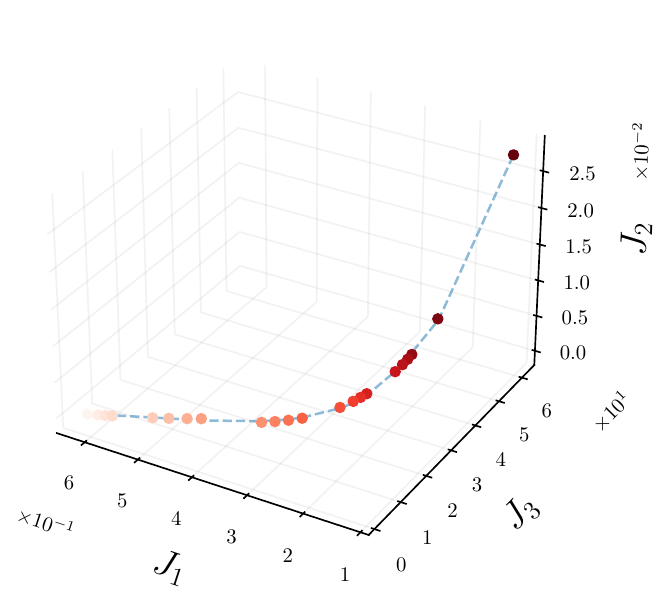}
    \caption{}
  \end{subfigure}
  \hfill
  \begin{subfigure}[b]{0.49\textwidth}
    \includegraphics[width=\textwidth]{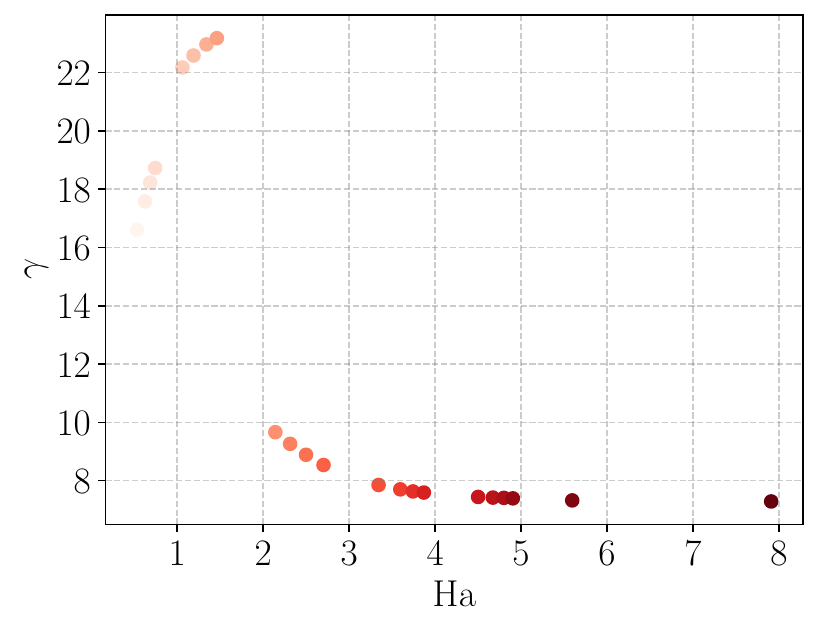}
    \caption{}
  \end{subfigure}
  \caption{Pareto's front in the (a) objective space and (b) the parameter space.}
  \label{pareto_front_2}
\end{figure}

\FloatBarrier
By running the multi-objective optimization from different initial conditions, we found the optimal parameters respecting the Karush-Kuhn-Tucker optimality condition. Figure \ref{pareto_front_2} shows the Pareto front collecting the optimal solutions in the objective space (a) and the parameter space (b). The markers with the same red shad in the two figures correspond to the same optimal solution. What stands out from these plots is the role of $\gamma$. For $\Ha<2$, an increase of $\Ha$ is accompanied by an increase of $\gamma$. A larger magnetic Gaussian compensates for the steepening of the interface due to $\Ha$, while at the same time keeping the Joule effect to a reasonably low value. For $\Ha>2$, an increase of $\Ha$ is accompanied by a decrease of $\gamma$. In this case, it is better to accept a steep film interface with a hump but with limited Joule heating. For larger, $\Ha$ the optimal $\gamma$ approach a plate at $\gamma\approx 8$. The Pareto front in Figure \ref{pareto_front_2} is tabulated in \ref{Tabella} for the reader convenience.

\begin{figure*}
\centering
  \begin{subfigure}[b]{0.49\textwidth}
    \includegraphics[width=\linewidth]{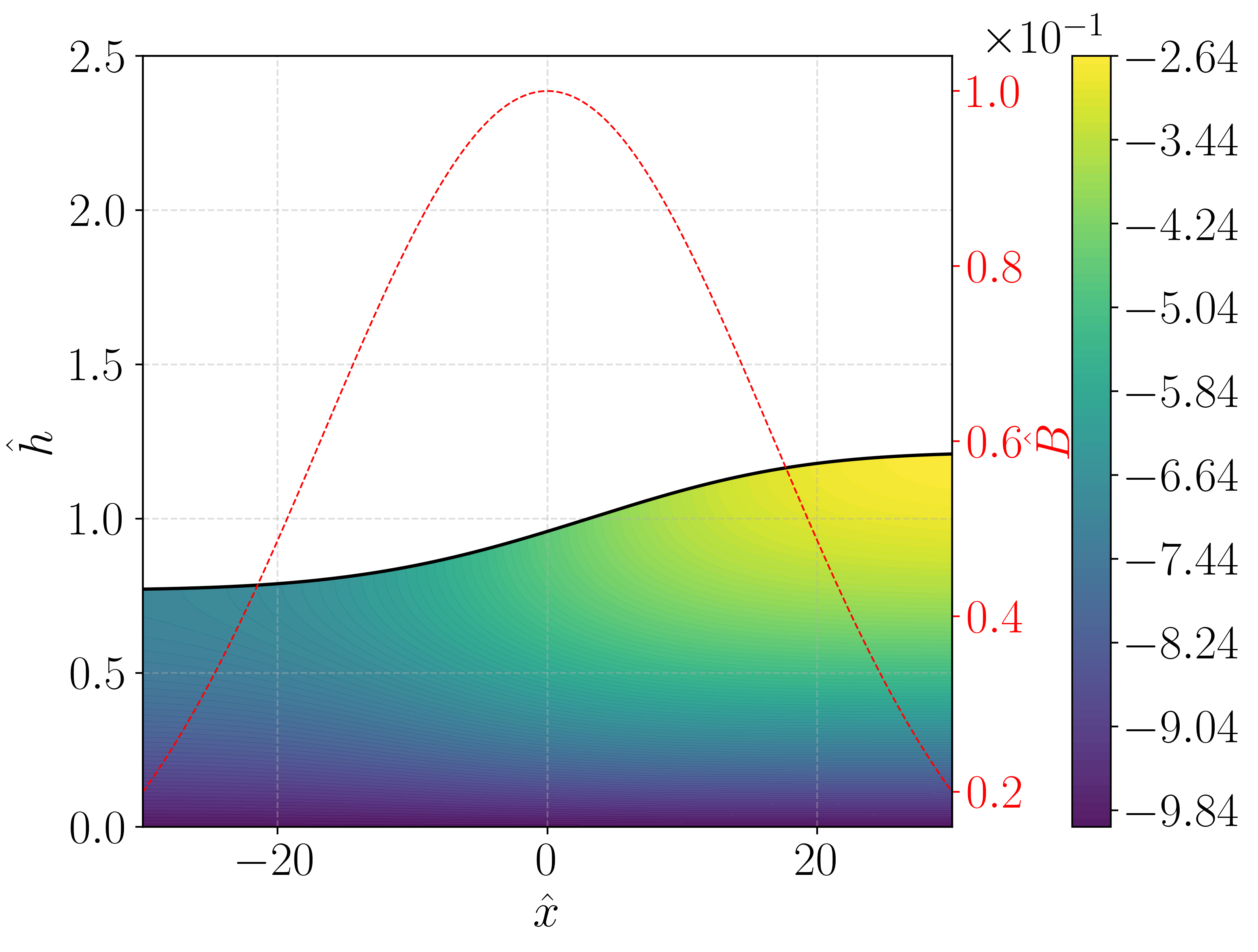}
    \caption{}
  \end{subfigure}
  \hfill
  \begin{subfigure}[b]{0.49\textwidth}
    \includegraphics[width=\linewidth]{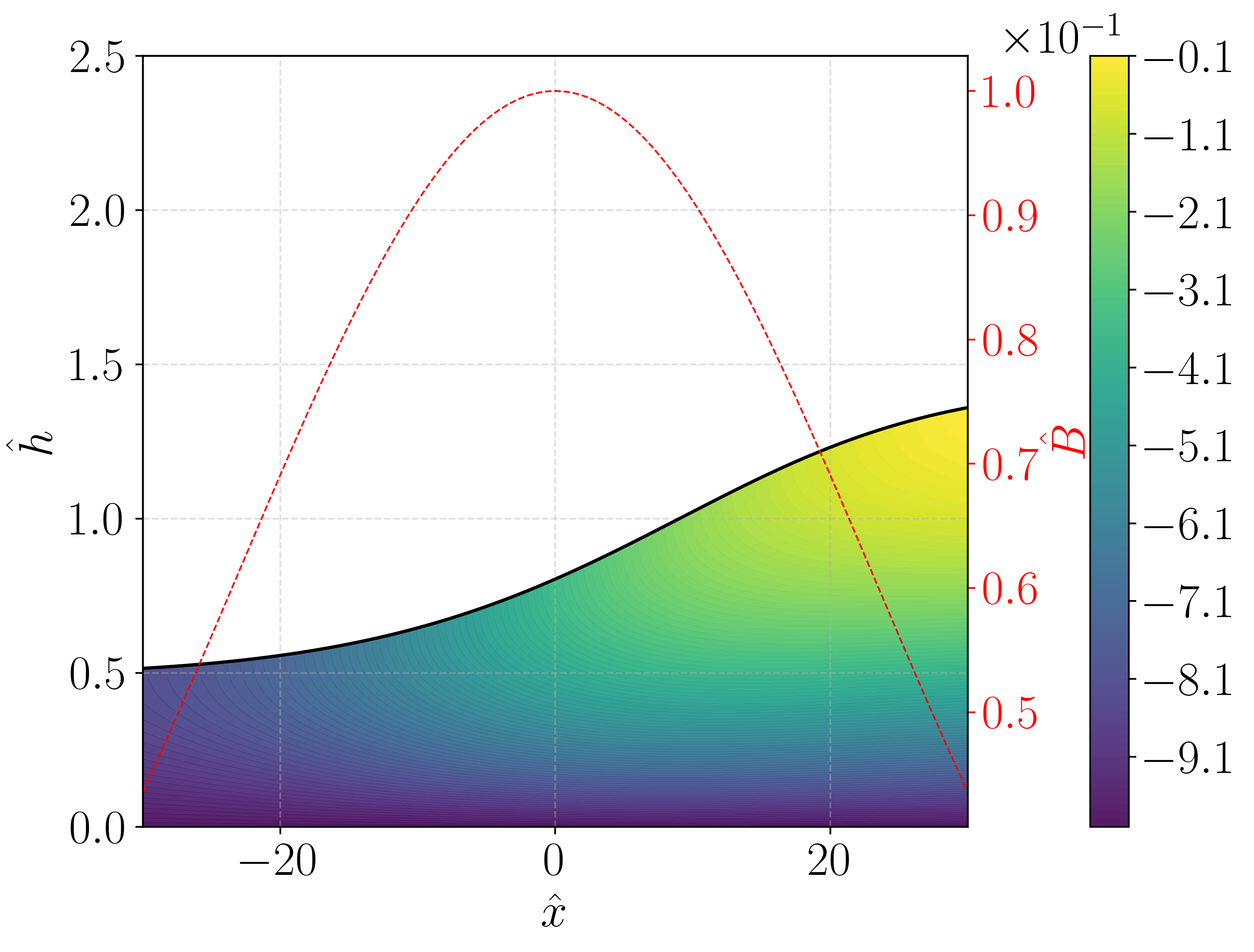}
    \caption{}
  \end{subfigure}
  \hfill
  \begin{subfigure}[b]{0.49\textwidth}
    \includegraphics[width=\linewidth]{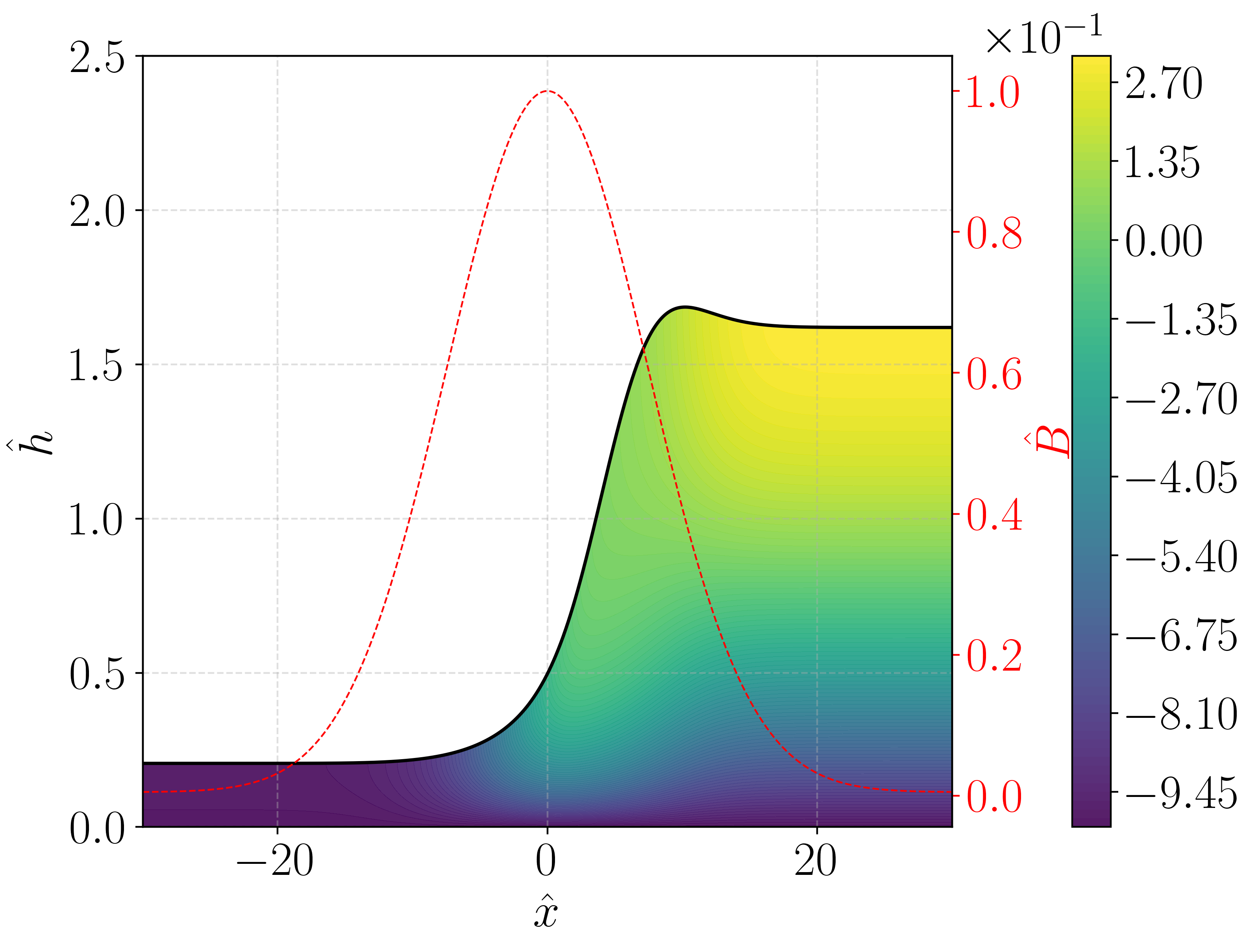}
    \caption{}
  \end{subfigure}
  \hfill
  \begin{subfigure}[b]{0.49\textwidth}
    \includegraphics[width=\linewidth]{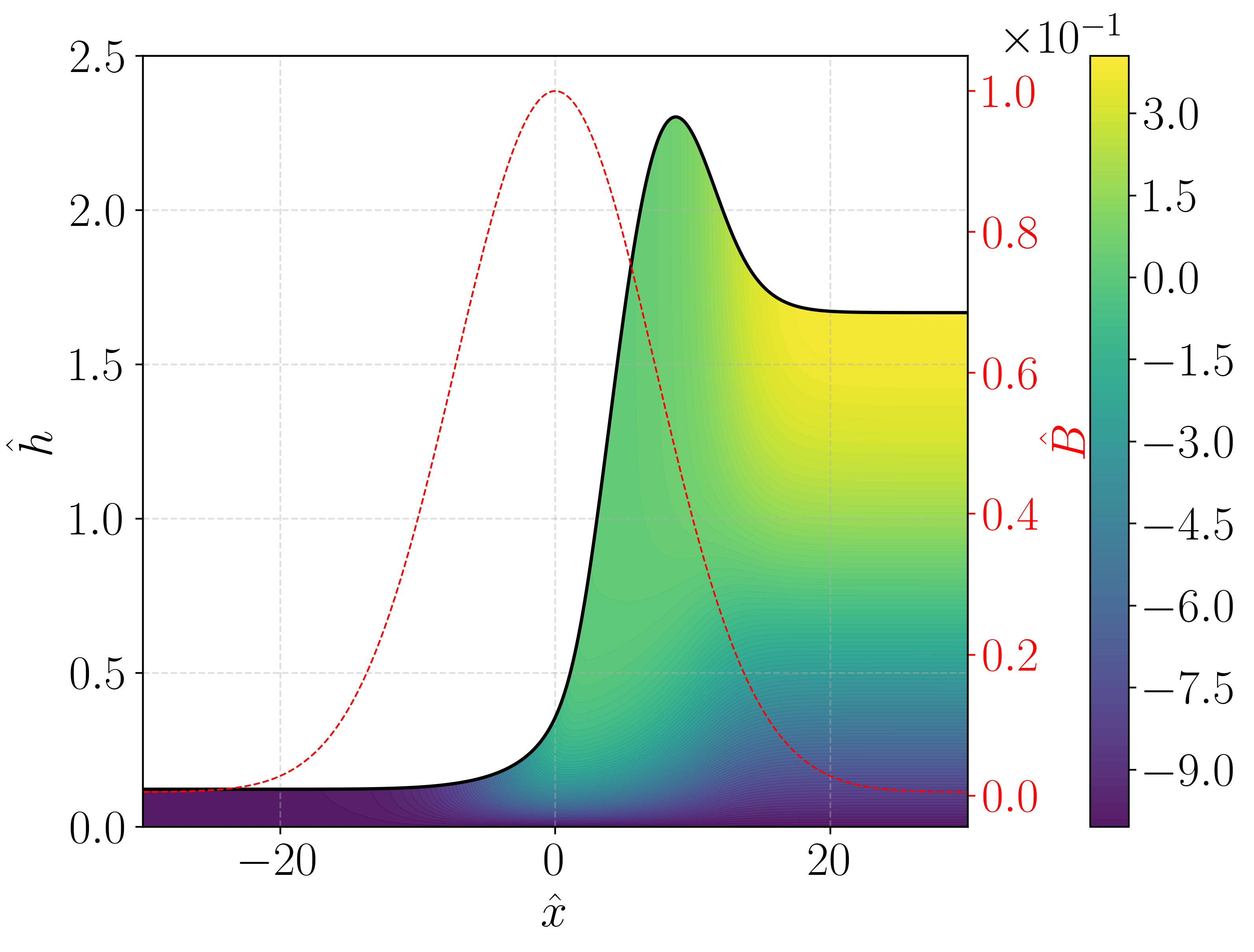}
    \caption{}
  \end{subfigure}
  \caption{Induced current colour plot of optimal solutions belonging to the Pareto front.}
  \label{different_pareto_optimal_solutions}
\end{figure*}

Figure~\ref{different_pareto_optimal_solutions} show a collection of solutions belonging to the Pareto front, with parameter values reported in Table \ref{tab:optimal_solutions}. As $\Ha$ increases, the hump starts to appear. Inside the hump, the current changes sing becoming positive. This means that the streamwise velocity changes sign in this region. This suggests the presence of a recirculation region under the hump, restricting the area where the flow can pass to the far-field part of the film. This creates an intense induced current around the magnetic peak with a sharp gradient. As a consequence, in this area and the far field, the liquid is strongly affected by an intense Joule heating.

\begin{table}[h]
\centering
\caption{Cost functions and parameters' values of four optimal solutions belonging to the Pareto front.}
\label{tab:optimal_solutions}
\begin{tabular}{cccccc}
Point & $J_1$ & $J_2$ & $J_3$    & $\Ha$ & $\gamma$ \\
\hline
(a)     & 0.62  & 2.15  & 2$\times 10^{-6}$  & 0.53  & 16.61    \\
(b)     & 0.45  & 11.56 & 4$\times 10^{-6}$ & 1.46  & 23.18    \\
(c)     & 0.2   & 33.42 & 0.004993 & 4.5   & 7.44     \\
(d)     & 0.12  & 61.17 & 0.027693 & 7.91  & 7.29    
\end{tabular}
\end{table}

\begin{figure}[h!]
\centering
\includegraphics[width=0.5\textwidth]{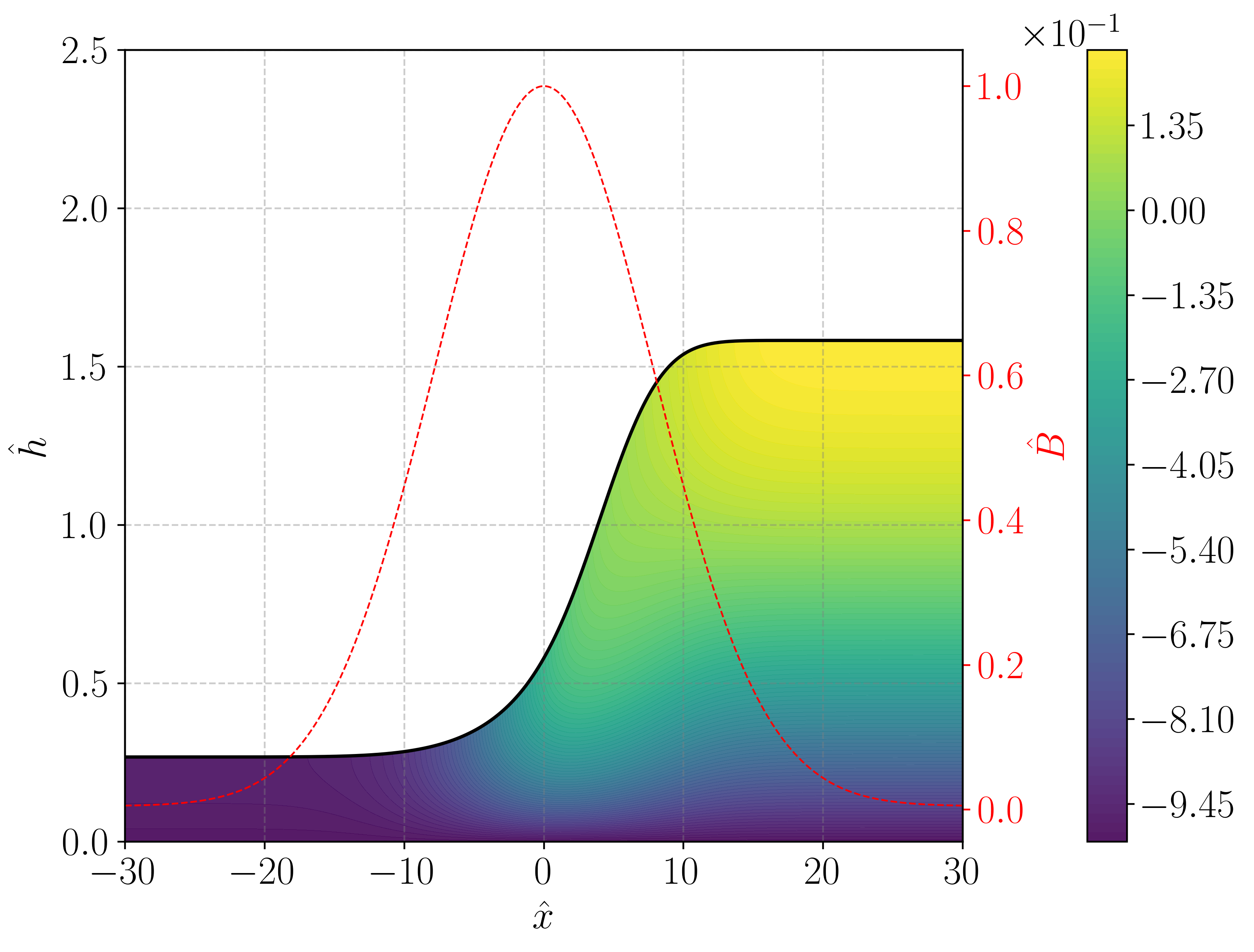}
\caption{Solution minimizing the sum of the three cost functions in terms of film thickness and current distribution for the multi-objective optimization problem.}
\label{optimal_solution}
\end{figure}

Figure~\ref{optimal_solution} shows the solutions that minimize the distance to the so-called ideal point, the origin of the normalized objective space reference system. The liquid film is smooth without any hump and with an intense but limited Joule heating, especially in the proximity of the moving strip. However, considering an operating velocity of $1[m/s]$, the dimensional final coating is at least one order of magnitude larger than the one required.

\FloatBarrier
\section{Conclusion}
\label{conclusion}

In this work, we presented a multi-objective optimization for electromagnetic wiping in galvannealing coating processing. The optimization balances the need to reduce the film thickness while preserving a smooth wiping meniscus and minimizing the Joule heating. The problem was solved via a multi-gradient descent optimization, providing the Pareto front in the objectives' and parameters' spaces.

What stands out from this analysis is that the liquid film overheating and the steepness of the wiping meniscus increase for $Ha>3$. Depending on the coating requirements, one can choose solutions favouring one objective over the other. This result could pave the way to a hybrid solution where the jet wiping is supported by magnetic wiping in the pre-wiping or a pre-heating phase before the galvannealing furnace. This would represent a good compromise between wiping efficiencies and an increase in the temperature of the liquid, which would reduce the effort of the galvannealing furnace in creating the Fe-Zn alloy.

\appendix

\section{Appendix A: Computation of the cost functions gradients}
\label{sec:gradients_computation}
To solve the optimization problem, we calculate the gradients of $J$ with respect to the magnetic parameters. The partial derivative of $J$ with respect to $\Ha$ is given by:

\begin{equation}
    \partial_{\Ha}J = \partial_{\Ha}J_1 + \partial_{\Ha}J_2 + \partial_{\Ha}J_3 
\end{equation}
where $\partial_{\Ha}J_{1}$ is easily obtained via \eqref{eq:gradient_flow_rate}:
\begin{equation}
    \partial_{\Ha}J_1 = \partial_{\Ha}\hat{q} = \hat{u}(\hat{h})\partial_{\Ha}\hat{h} + \int_{0}^{\hat{h}}\,\partial_{\Ha}\hat{u}(\hat{y})\,d\hat{y}\,,
\end{equation}

The partial derivative $\partial_{\Ha}J_{2}$ is given by the expression: 
\begin{equation}
\partial_{\Ha}J_2 = \int_{x_{1}}^{x_{2}}\,\partial_{\hat{x}\hat{x}}\hat{h}\partial_{\hat{x}\hat{x}}(\partial_{\Ha}\hat{h})\,d\hat{x},
\end{equation} having considered a domain $x\in[x_1,x_2]$

The partial derivative of the Joule heating term $\partial_{\Ha}J_{3}$ is computed using the current density sensitivity $\partial_{\Ha}\hat{j}_z$: 

\begin{equation}
    \begin{split}
\partial_{\Ha}J_3 =& \frac{1}{2}\int_{x_{1}}^{x_{2}}\,\Ha^2\hat{j}(\hat{h})^2\partial_{\Ha}\hat{h}\,d\hat{x} +\int_{x_{1}}^{x_{2}}\int_{0}^{\hat{h}}\, \Ha\hat{j}(\hat{y})^2\,d\hat{y}d\hat{x} +\int_{x_1}^{x_{2}}\int_{0}^{\hat{h}}\Ha^2\hat{j}\partial_{\Ha}\hat{j}\,d\hat{x}.
    \end{split}
\end{equation} 

The gradient of $J$ with respect to $\gamma$ is given by:
\begin{equation}
    \partial_{\gamma}J = \partial_{\gamma}J_1 + \partial_{\gamma}J_2 + \partial_{\gamma}J_3,
\end{equation}
where:
\begin{subequations}
\begin{equation}
    \partial_{\gamma}J_1 = \partial_{\gamma}\hat{q} = \hat{u}(\hat{h})\partial_{\hat{B}}\hat{h}\partial_{\gamma}\hat{B} + \int_{0}^{\hat{h}}\,\partial_{\hat{B}}\hat{u}(\hat{y})\partial_{\gamma}\hat{B}\,d\hat{y}\,,
\end{equation}
\begin{equation}
    \partial_{\gamma}J_2 =
    \int_{x_{1}}^{x_{2}}\,\partial_{\hat{x}\hat{x}}\hat{h}\partial_{\hat{x}\hat{x}}(\partial_{\hat{B}}\hat{h})\partial_{\gamma}\hat{B}\,d\hat{x},
\end{equation}
\begin{equation}
    \begin{split}
        \partial_{\gamma}J_3 =& \frac{1}{2}
        \int_{x_{1}}^{x_{2}}\,\Ha^2\hat{j}(\hat{h})^2\partial_{\hat{B}}\hat{h}\partial_{\gamma}\hat{B}\,d\hat{x}+\int_{x_{1}}^{x_{2}}\int_{0}^{\hat{h}}\,
        \Ha^2\hat{j}\partial_{\hat{B}}\hat{j}\partial_{\gamma}\hat{B}\,d\hat{y}d\hat{x}.
    \end{split}
\end{equation}
\end{subequations}

\section{Appendix B: Tabulated Values of the identified Pareto front}

Table \ref{Tabella} tabulates the identified Pareto front shown in Figure \ref{pareto_front_2}.

\begin{table}[ht!]
\centering
\caption{Value of the magnetic parameters ($Ha,\gamma$) and the objective functions ($J_1,J_2,J_3$) along the identified Pareto front.}
\label{Tabella}
\begin{tabular}{ccccc}
Ha    & $\gamma$ & $J_1$ & $J_2$     & $J_3$  \\ \hline
0.53, & 16.61,   & 0.62, & 2e-06,    & 2.15,  \\
0.63, & 17.58,   & 0.6,  & 2e-06,    & 2.9,   \\
0.69, & 18.23,   & 0.59, & 2e-06,    & 3.43,  \\
0.74, & 18.73,   & 0.58, & 2e-06,    & 3.96,  \\
1.06, & 22.18,   & 0.52, & 2e-06,    & 7.21,  \\
1.19, & 22.59,   & 0.5,  & 2e-06,    & 8.59,  \\
1.34, & 22.97,   & 0.47, & 3e-06,    & 10.23, \\
1.46, & 23.18,   & 0.45, & 4e-06,    & 11.56, \\
2.14, & 9.66,    & 0.36, & 0.000261, & 15.61, \\
2.31, & 9.26,    & 0.34, & 0.000368, & 16.88, \\
2.5,  & 8.89,    & 0.33, & 0.000518, & 18.23, \\
2.7,  & 8.54,    & 0.31, & 0.000733, & 19.68, \\
3.34, & 7.85,    & 0.26, & 0.00174,  & 24.34, \\
3.6,  & 7.7,     & 0.25, & 0.002283, & 26.25, \\
3.74, & 7.63,    & 0.24, & 0.002646, & 27.38, \\
3.87, & 7.59,    & 0.23, & 0.002965, & 28.38, \\
4.5,  & 7.44,    & 0.2,  & 0.004993, & 33.42, \\
4.67, & 7.42,    & 0.2,  & 0.005651, & 34.81, \\
4.8,  & 7.41,    & 0.19, & 0.006165, & 35.87, \\
4.9,  & 7.39,    & 0.19, & 0.006634, & 36.7,  \\
5.6,  & 7.32,    & 0.17, & 0.010218, & 42.26, \\
7.91  & 7.29     & 0.12  & 0.027693  & 61.17 
\end{tabular}
\end{table}

% \begin{contributions}
% \textbf{Manuel Ratz}: Methodology, Software, Validation, Experiments, Formal analysis, Investigation, Data Curation, Writing - Original Draft, Visualisation. \textbf{Miguel A. Mendez}: Conceptualisation, Methodology, Software, Investigation, Resources, Writing - Review \& Editing, Supervision, Project Administration, Funding Acquisition.
% \end{contributions}

\begin{acknowledgements}
F.Pino is supported by an F.R.S.-FNRS FRIA grant, and his PhD research project is funded by Arcelor-Mittal. B.Scheid is Research Director at F.R.S.-FNRS.
\end{acknowledgements}
% \section*{Declarations}

% \begin{dataavail}
% Data sets generated during the current study are available from the corresponding author on reasonable request.
% \end{dataavail}

% \begin{ethics}
% Not applicable.
% \end{ethics}

% \begin{conflicts}
% The authors declare no competing interests.
% \end{conflicts}

\bibliographystyle{spbasic}
\bibliography{Main.bib}

\begin{thebibliography}{22}
\providecommand{\natexlab}[1]{#1}
\providecommand{\url}[1]{{#1}}
\providecommand{\urlprefix}{URL }
\expandafter\ifx\csname urlstyle\endcsname\relax
  \providecommand{\doi}[1]{DOI~\discretionary{}{}{}#1}\else
  \providecommand{\doi}{DOI~\discretionary{}{}{}\begingroup \urlstyle{rm}\Url}\fi
\providecommand{\eprint}[2][]{\url{#2}}

\bibitem[{Barreiro-Villaverde et~al.(2021)Barreiro-Villaverde, Gosset, and Mendez}]{Barreiro-Villaverde2021}
Barreiro-Villaverde D, Gosset A, Mendez M (2021) {On the dynamics of jet wiping: Numerical simulations and modal analysis}. Physics of Fluids 33(6), \doi{10.1063/5.0051451}

\bibitem[{Bregand et~al.(2010)Bregand, Crahay, and Gerkens}]{european_union_report}
Bregand O, Crahay J, Gerkens P (2010) New wiping techniques to efficiently produce suitable coating layers at high speed in the hot dip galvanising process. \doi{10.2777/86670}

\bibitem[{Calver and Enright(2017)}]{calver2017numerical}
Calver J, Enright W (2017) Numerical methods for computing sensitivities for odes and ddes. Numerical Algorithms 74(4):1101--1117

\bibitem[{Center(2014)}]{galvingo_galvanneal}
Center G (2014) 1.3 galvanneal – differences from galvanize. Tech. rep., GalvInfo Notes, International Zinc Association

\bibitem[{D{\'e}sid{\'e}ri(2012)}]{desideri2012multiple}
D{\'e}sid{\'e}ri JA (2012) Multiple-gradient descent algorithm (mgda) for multiobjective optimization. Comptes Rendus Mathematique 350(5-6):313--318

\bibitem[{Dumont et~al.(2011)Dumont, Ernst, Fautrelle, Grenier, Hardy, and Anderhuber}]{Dumont2011d}
Dumont M, Ernst R, Fautrelle Y, Grenier B, Hardy JJ, Anderhuber M (2011) {New DC electromagnetic wiping system for hot-dip coating}. COMPEL - The International Journal for Computation and Mathematics in Electrical and Electronic Engineering 30(5):1663--1671, \doi{10.1108/03321641111152810}

\bibitem[{Ernst et~al.(2009)Ernst, Fautrelle, Bianchi, and Iliescu}]{ernst2009working}
Ernst R, Fautrelle Y, Bianchi A, Iliescu M (2009) Working principle of an electromagnetic wiping system. Magnetohydrodynamics 45(1):59--71

\bibitem[{Giordanengo et~al.(2000)Giordanengo, Moussa, Makradi, Chaaba, and Gasser}]{giordanengo2000new}
Giordanengo B, Moussa AB, Makradi A, Chaaba H, Gasser J (2000) New precise determination of the high temperature unusual temperature dependent thermopower of liquid divalent cadmium and zinc. Journal of Physics: Condensed Matter 12(15):3595

\bibitem[{Goodwin(2013)}]{goodwin2013developments}
Goodwin FE (2013) Developments in the production of galvannealed steel for automotive. Transactions of the Indian Institute of Metals 66(5):671--676

\bibitem[{Gosset and Buchlin(2006)}]{gosset2007jet}
Gosset A, Buchlin JM (2006) Jet wiping in hot-dip galvanization. Journal of Fluids Engineering 129(4):466--475, \doi{10.1115/1.2436585}

\bibitem[{Gosset et~al.(2019)Gosset, Mendez, and Buchlin}]{Gosset2019}
Gosset A, Mendez MA, Buchlin JM (2019) {An experimental analysis of the stability of the jet wiping process: Part I – Characterization of the coating uniformity}. Experimental Thermal and Fluid Science 103:51--65, \doi{10.1016/j.expthermflusci.2018.12.029}

\bibitem[{Hocking et~al.(2011)Hocking, Sweatman, Fitt, and Breward}]{hocking2011deformations}
Hocking GC, Sweatman W, Fitt A, Breward C (2011) Deformations during jet-stripping in the galvanizing process. Journal of Engineering Mathematics 70:297--306

\bibitem[{Kalliadasis et~al.(2011)Kalliadasis, Ruyer-Quil, Scheid, and Velarde}]{kalliadasis2011falling}
Kalliadasis S, Ruyer-Quil C, Scheid B, Velarde MG (2011) Falling liquid films, vol 176. Springer Science \& Business Media

\bibitem[{Kistler and Schweizer(1997)}]{kistler1997coating}
Kistler SF, Schweizer PM (1997) Coating science and technology: An overview. Liquid Film Coating pp 3--15

\bibitem[{{Lloyd-Jones, C and Barker, HA and Worner}(1998)}]{Lloyd-JonesCandBarkerHAandWorner1998d}
{Lloyd-Jones, C and Barker, HA and Worner} V (1998) {Investigation into magnetic wiping techniques as alternative to gas wiping on hot dip galvanising lines}. Ironmaking {\&} steelmaking 25(2):117

\bibitem[{Mao et~al.(2008)Mao, Aleksandrova, and Molokov}]{mao2008joule}
Mao J, Aleksandrova S, Molokov S (2008) Joule heating in magnetohydrodynamic flows in channels with thin conducting walls. International journal of heat and mass transfer 51(17-18):4392--4399

\bibitem[{Mendez et~al.(2019)Mendez, Gosset, and Buchlin}]{mendez2019experimental}
Mendez M, Gosset A, Buchlin JM (2019) Experimental analysis of the stability of the jet wiping process, part ii: Multiscale modal analysis of the gas jet-liquid film interaction. Experimental Thermal and Fluid Science 106:48--67

\bibitem[{Mendez et~al.(2021)Mendez, Gosset, Scheid, Balabane, and Buchlin}]{mendez2021dynamics}
Mendez M, Gosset A, Scheid B, Balabane M, Buchlin JM (2021) Dynamics of the jet wiping process via integral models. Journal of Fluid Mechanics 911

\bibitem[{Milojkovic et~al.(2019)Milojkovic, Antognini, Bergamin, Faltings, and Musat}]{milojkovic2019multi}
Milojkovic N, Antognini D, Bergamin G, Faltings B, Musat C (2019) Multi-gradient descent for multi-objective recommender systems. arXiv preprint arXiv:200100846

\bibitem[{Smolentsev et~al.(1997)Smolentsev, Tananaev, and Dajeh}]{smolentsev1997laminar}
Smolentsev SY, Tananaev A, Dajeh D (1997) Laminar heat transfer in the mhd flow in a rectangular duct 1. numerical and analytical investigations. Magnetohydrodynamics c/c of Magnitnaya Gidrodinamika 33:148--154

\bibitem[{Thornton and Graff(1976)}]{thornton1976analytical}
Thornton JA, Graff HF (1976) An analytical description of the jet finishing process for hot-dip metallic coatings on strip. Metallurgical Transactions B 7(4):607--618

\bibitem[{Tuck(1983)}]{tuck1983continuous}
Tuck E (1983) Continuous coating with gravity and jet stripping. The Physics of fluids 26(9):2352--2358

\end{thebibliography}

% \renewcommand{\thesection}{Appendix \Alph{section}}
% \setcounter{section}{0}
% \renewcommand\theequation{A\arabic{equation}}
% \setcounter{equation}{0}

% \section*{Appendix A}
% \section{RBF PUM regression in 3D}
% \label{app:pum_regression_3d}

% \section*{Appendix B}
% \section{Reynolds stress computation in 3D}

\end{document}